\documentstyle[12pt,epsf]{article}

\begin{document}
\title{\bf Simple Applications of  Effective Field Theory and Similarity
Renormalization Group Methods }
\author{S\'ergio Szpigel and Robert J. Perry
\\ \\{ \it Department of Physics}\\ { \it The Ohio State University,
Columbus, OH 43210}\\}    

\date{\today}
\maketitle
\abstract{
We use two renormalization techniques, Effective Field Theory and the
Similarity Renormalization Group, to solve simple Schr{\"o}dinger
equations with delta-function potentials in one and two dimensions. The
familiar one-dimensional delta-function does not require renormalization, but
it provides the simplest example of a local interaction that can be replaced
by a sequence of effective cutoff interactions that produce controllable
power-law errors. The two-dimensional delta-function leads to logarithmic
divergences, dimensional transmutation and asymptotic freedom, providing
an example of some of the most important renormalization problems in
gauge field theories. We concentrate on the power-law analysis of errors
in low-energy observables. The power-law suppression of the effects of
irrelevant operators is critical to the success of field theory, and
understanding them turns renormalization group techniques into powerful
predictive tools for complicated problems where exact solutions are not
available.
}

\vfill
\eject
\vskip .25in
%

\section{Introduction}
The theory of renormalization is one of the greatest successes of physics in
the twentieth century~\cite{brown}. Early attempts to marry quantum mechanics
with special relativity naturally led to the use of local interactions,
maintaining the causal paradigm that has driven physics for centuries while
avoiding signals that propagate at speeds greater than that of light.
Early encounters with perturbative divergences led some of the best theorists
in the world to question the foundations of quantum mechanics, and eventual
successes at fitting precise atomic experimental data led to the universal
acceptance of renormalization recipes that were acknowledged to make little
sense~\cite{schwinger}. Initially the perturbative renormalization of QED
required theorists to match perturbative expansions in powers of a bare and
physical electronic charge~\cite{dirac1}, but the bare charge clearly diverges
logarithmically in QED and the success of an expansion in powers of such a
coupling was mysterious at best~\cite{dirac3}.

The first steps towards making sense of renormalization theory were taken
in the 1950's with the invention of the perturbative renormalization group
~\cite{stueck,gellmannlow,landaupole,bogol1,bogol}, although serious
investigators found the theory was still plagued by non-convergent sums
because QED is not asymptotically free. The development of Wilson's
renormalization group formalism~\cite{wilson9,wilson10} and the discovery of
asymptotic freedom~\cite{free} allowed physicists to produce a logically
reasonable picture of renormalization in which perturbative expansions at any
high energy scale can be matched with one another, with no necessity to deal
with intermediate expansions in powers of a large parameter. But in all
theories of interest the running coupling becomes large at some scale and the
lattice~\cite{wilsonlattice} provides a tool for attacking gauge theories
non-perturbatively. This shifted the problems to achieving a continuum limit
in which the lattice spacing can be taken to zero implicitly and set the stage
for the development of a standard model that rests heavily on asymptotically
free field theories.

Although this story took a dramatic turn with the invention of string
theories, its main themes survive in essentially all predictive field theory
calculations and the control the renormalization group provides has made it
possible for physicists to accept the standard model as an effective theory
that is accurate over a very large range of scales that covers all currently
experimentally accessible energies. It has also led to the acceptance of the
development of effective field theories as a fundamental endeavor that goes
far beyond the apparently arbitrary introduction of non-local potentials,
which has persistently undermined broad interest in attempts to model
complicated interactions such as the strong nuclear interaction.

We make no attempt to survey this history, nor are we able to provide a
complete review of the two methods illustrated in this article. We take
advantage of the fact that the divergences that provided the original impetus
leading to renormalization theory result entirely from the use of local
interactions. To understand the most important aspects of renormalization
theory requires only a background in nonrelativistic quantum mechanics,
because as has been long known the divergences of field theory are directly
encountered when one tries to impose locality on the Schr{\"o}dinger equation.
In this case the only available interactions that are local at all scales are
delta functions and derivatives of delta functions. These divergences can be
regulated by the introduction of a cutoff, and the artificial effects of this
cutoff must be removed by renormalization. The simplicity of the one-body
Schr{\"o}dinger equation makes it possible to renormalize the theory exactly,
disentangling the effects of locality from the complicated many-body effects
and symmetry constraints encountered in realistic field theories. There is a
large literature on the subject~\cite{zeldo}-\cite{phillips}, largely
pedagogical.

Effective Field Theory (EFT) is primarily used to replace complicated
`fundamental' theories such as QCD with simpler theories such as chiral
perturbation theory~\cite{chpt}. These simpler effective theories are to be
used only at `low' energy scales at which the fundamental degrees of freedom
and interactions cannot be resolved, and the scaling analysis that we
illustrate makes this limitation clear. EFT ~\cite{weinberg}-\cite{kolck} is
motivated by Wilson's work on the renormalization
group~\cite{wilson9,wilson10}; however, practitioners typically employ only
basic renormalization group tools. 

EFT and essentially all renormalization techniques rely on the assumption
that physics at a low energy scale and/or long distance scale is
insensitive to the details of the underlying physics at a high energy
scale and/or short distance scale. Microscopic degrees of freedom and
their interactions can be replaced by effective macroscopic degrees of
freedom and their effective interactions. Moreover, these effective
degrees of freedom can be treated as if they were point-like and their
effective interactions are local at the scale of interest. These
assumptions have clearly been critical through the entire evolution of
physics. However, the detailed procedure for producing effective
interactions and the understanding of how errors depend on the parameters
in these interactions are relatively new.

One of the most interesting recent applications of EFT is to nuclear physics
and the two-nucleon problem in particular ~\cite{ordonez92}-\cite{cohen99c}.
Traditionally nuclear theorists have developed nonlocal two-body interactions,
sometimes motivated by meson exchange models, that are adjusted to fit
scattering phase shifts and properties of the deuteron. These nonlocal
interactions are typically parameterized by a few masses and couplings, but
there has been no systematic, logically simple procedure for improving
interactions and many nonlocal interactions work equally well. In EFT the
cutoff is regarded as artificial and parameters are allowed to run with the
cutoff to remove its effects as much as possible. There is a systematic
procedure for improving the effective interactions by adding operators, and
the goal of relating the complicated nucleon-nucleon interaction to QCD
may become much more realistic.

Recent work to extend this approach to deal with the three-nucleon
problem ~\cite{bedaque99a,bedaque99b} and to explicitly include mesons has
uncovered problems that go beyond a straightforward application of EFT. The
fact that the two-nucleon interaction is an irrelevant operator, but the
scattering phases shifts are large, has led to an understanding that naive
dimensional scaling is violated. There is a non-free fixed point that contains
this interaction and presumably alters the scaling properties of all operators
~\cite{richardson97,birse98a,birse98b,birse98c}. Condensed matter theorists
have long appreciated the importance of understanding the scaling behavior of
interactions around non-Gaussian fixed points, but there are few field theory
examples that are relevant to nuclear and particle physics. The study of
effective nuclear field theories may offer such examples and then drive
interesting new developments in EFT. 

The similarity renormalization group (SRG) is a very recent development
invented by Stan G{\l}azek and Ken Wilson~\cite{wilgla1,wilgla2}, and
independently by Franz Wegner~\cite{wegner}. Most of the
applications of the SRG are in light-front field theory
~\cite{LFQCD}-\cite{gubankova99}, but there are a growing number of
applications in other areas ~\cite{szczepaniak96}-\cite{glazek99}.

In EFT a cutoff is introduced along with a fixed number of effective
interactions. The resultant hamiltonian is used to compute low energy
observables, and the strengths ({\it i.e.}, couplings) of the effective
interactions are adjusted to fit a subset of the low energy data. It is
sometimes possible to explicitly derive equations that govern how these
couplings evolve as the cutoff changes, an approach that parallels the
Gell-Mann--Low approach to the renormalization group~\cite{gellmannlow}, and
by studying these equations the scaling behavior of these couplings can be
understood. In the SRG, as in Wilson's original renormalization group
formalism~\cite{wilson65,wilson70}, transformations that explicitly run the
cutoff are developed. These transformations are the group elements that give
the renormalization group its name.

In his earliest work ~\cite{wilson65,wilson70} Wilson exploited a
transformation originally invented by Claude Bloch~\cite{bloch}. It uses a
cutoff on the states themselves, and as the cutoff is lowered, states are
removed from the Hilbert space. If the hamiltonian is viewed as a matrix,
these cutoffs can be seen as limiting the size of this matrix and the
transformation reduces this size, as illustrated in Fig. 1a. Wilson introduced
a rescaling operation to allow transformed hamiltonians to be compared with
initial hamiltonians, despite the fact that they act in different spaces;
however, the Bloch transformation is ill-defined and even in perturbation
theory it leads to artificial divergences. These divergences come from the
small energy differences between states retained and states removed by the
transformation, and they appear in the form of small energy denominators in
the perturbative expansion of the transformed hamiltonian. These small energy
denominator problems led Wilson to abandon the hamiltonian formulation of field
theory in favor of path integral formulations, but the virtues of the
hamiltonian formulation over the path integral formulation for many
problems remains.

\begin{figure}
\centerline{\epsffile{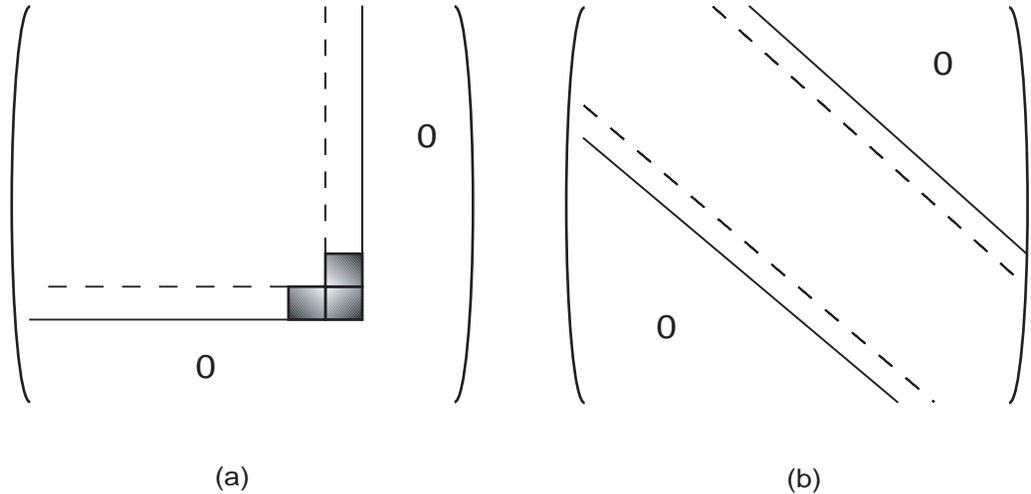}}
\vspace{0.5cm}
\caption{Two ways to run a cutoff on free energy.  In (a) a cutoff
on the magnitude of the energy is lowered from the solid to the
dashed lines, with problems resulting from the removed shaded
region. In (b) a cutoff on how far off diagonal matrix elements
appear is lowered from the dashed to the solid lines.}
\end{figure}

The breakthrough provided by the SRG is that the transformations are
typically unitary, making them well-defined, and they run a cutoff on
energy differences rather than on individual states, as illustrated in Fig. 1b.
Again viewing the hamiltonian as a large matrix, these cutoffs limit the
off-diagonal matrix elements and as they are reduced the hamiltonian is forced
towards diagonal form. The perturbative expansion for transformed hamiltonians
contains no small energy denominators, so the expansion breaks down only
when interactions become sufficiently strong, in which case perturbation
theory should fail in any case.

Although the SRG has not yet been applied to a wide range of problems, it may
be an important new tool both for attacking field theories and
non-relativistic many-body problems. As we will illustrate, the effective
interactions that arise in the SRG are nearly identical to those
encountered in EFT; however, the existence and use of an explicit
transformation that lowers the cutoff allows one to directly study
scaling behavior before using the hamiltonian to fit data. 

When the SRG is used with coupling coherence ~\cite{ochme,coupcoh}, which we
explain in the text, it allows one to construct effective theories with
the same number of free parameters as the underlying `fundamental' theory. For
the delta-function examples there is one fundamental parameter, the strength
of the regulated delta-function as the cutoff is removed. In EFT this
single parameter is replaced by an arbitrarily large number of effective
parameters that are fixed by data. In the SRG with coupling coherence,
there is still one fundamental coupling and all new couplings are
fixed perturbative functions of the fundamental coupling. It is the
renormalization group flow of the added couplings, and a boundary condition
that they vanish when the fundamental coupling is taken to zero, that fixes
their dependence on the fundamental coupling.

The examples we use in this article do not illustrate non-Gaussian fixed
points, so their scaling properties are driven by naive dimensional
analysis. However, we will see that even in these cases scaling behavior
of effective hamiltonians derived using a perturbative similarity
renormalization group can be very complicated. We will see that errors in
EFT scale as inverse powers of the cutoff, which is what one expects
since these operators are irrelevant. However, in the perturbative SRG
there are additional errors arising from the approximate treatment of the
fundamental running coupling and the approximate treatment of the
relation between this coupling and the new couplings of irrelevant
operators. 

The easiest way to understand errors in the perturbative SRG
is by comparison with EFT. The same set of operators are found in the SRG
hamiltonian as in the EFT hamiltonian, but the coefficients are
approximated using the perturbative SRG equations instead of being fit to
data.

In a realistic calculation the marginal coupling, which corresponds
to the strength of the regulated delta function, would be fit to data. In
order to clearly illustrate the logarithmic errors that result from using
the perturbative SRG equations, we approximate this marginal coupling in this
article rather than renormalizing it nonperturbatively by fitting data. The
strengths of the irrelevant operators, which correspond to derivatives of
the regulated delta function, are approximated using expansions in powers
of the approximate running coupling that are fixed by coupling coherence.
The approximate running coupling differs from the exact running coupling
by inverse powers of logarithms of the cutoff, and the error analysis for
the binding energy displays the resultant inverse logarithmic errors in
addition to power-law errors similar to those seen in EFT. In addition there
are errors in the strengths of the irrelevant operators resulting from using
a truncated expansion in powers of the running coupling and an
approximate running coupling, both of which introduce inverse
logarithmic errors in addition to the power-law errors seen for these
operators in EFT.

We do not argue that EFT or that SRG is to be preferred. Both allow us to
produce effective hamiltonians whose errors can be understood and thereby
controlled. In some problems it will be clear that it is much easier to
produce an approximate hamiltonian using EFT, but this requires the
calculation of scattering observables and in field theories it may require the
use of much of the existing data for a system just to fix the parameters. In
the SRG the number of parameters equals the number of parameters in the
underlying `fundamental' theory, but the SRG equations become extremely
complicated as one goes beyond second order. In addition EFT can be used even
when there is no underlying theory with the same degrees of freedom, but
coupling coherence should fail in this case. This issue deserves investigation.

There are interesting questions about whether the SRG formalism can be
used with coupling coherence when there is no clear connection between the
effective theory and any underlying fundamental theory. For example, the
number of couplings in chiral perturbation theory far exceeds the number
of parameters in QCD. If chiral effective theories are exact representations
of QCD in the low energy limit, coupling coherence would lead one to expect
that the number of independent couplings may be the same, so that there
must be constraints on the coupling in chiral field theories that have not been
determined. This article does not explore some of the most interesting
questions, but it is intended to clearly lay the groundwork required to
initiate such investigations.

%

%

\section{Effective Field Theory Approach}

EFT provides a systematic procedure for improving the approximate description 
of composite effective degrees of freedom and their dynamics by increasing the 
number of local effective interactions. The effects of high-energy 
(short-distance) degrees of freedom are incorporated in effective interactions 
whose strength is adjusted to fit appropriate low-energy (long-distance) 
results. The keys to maintaining predictive power in EFT are approximate 
locality and an expansion in powers of a small ratio of low-to-high energy 
scales or equivalently, short-to-long distance scales.

We begin by discussing the formulation of an effective description for a system 
of two non-relativistic particles that interact via a Dirac delta-function 
potential in D-dimensions, following Lepage's presentation of effective field
theory ~\cite{Lepage}. As mentioned in the Introduction, there is a large
literature on delta-function interactions where all of the basic results we
need can be found ~\cite{zeldo}-\cite{phillips}. The 
hamiltonian describing the relative motion of the system is given by 
\begin{equation}
H=-{\bf \nabla}_{\bf r}^{2}-\alpha_0 \; \delta^{(D)}({\bf r}) \; .
\end{equation}
\noindent
We use units in which $\hbar=1$ and choose the coupling $\alpha_0$ to be 
dimensionless. In the $D=1$ case the kinetic energy operator is implicitly 
divided by $2m=1$, where $m$ is the two-particle system reduced mass. In the 
$D=2$ case a mass scale is not necessary, since with a dimensionless coupling 
the hamiltonian is scale invariant.

First, we define a regularized delta-function potential using a cutoff 
$\lambda$ that  eliminates contributions from processes involving  momentum 
transfer $|{\bf q}| >\lambda$ (short-distance dynamics):
\begin{equation}
\delta^{(D)}_{\lambda}({\bf r})=\frac{1}{(\sqrt{2\pi}}\; \lambda)^D
\;  e^{-\frac{\lambda^2 \; r^2}{2}} \; .
\end{equation}
\noindent
The cutoff also regulates the interaction at $r=0$, avoiding possible 
divergences in the calculation of matrix elements.
\noindent
By Fourier transforming we obtain 
\begin{equation}
{\tilde \delta}^{(D)}_{\lambda}({\bf q})=\frac{1}{(2\pi)^D}\; e^{-\frac{q^2}{2 
{\lambda}^2}} \; ,
\end{equation}
\noindent
where ${\bf q}={\bf p}-{\bf p'}$ is the transferred momentum.

Next, we introduce an effective potential that consists of a series of 
approximately local contact interactions
\begin{eqnarray}
V_{\lambda}({\bf r},{\bf r'})&=&\delta^{(D)}({\bf r}-{\bf r'})\; 
\left[C_0(\lambda) \; \delta^{(D)}_{\lambda}({\bf r})+ \frac{1}{2\lambda^2}\; 
C_2(\lambda) \; {\bf \nabla}_{\bf r}^{2}\delta^{(D)}_{\lambda}({\bf r})\right. 
\nonumber\\
&+&\left.\frac{1}{4\lambda^4}\; C_4(\lambda) \; {\bf \nabla}_{\bf 
r}^{4}\delta^{(D)}_{\lambda}({\bf r})+ \dots \right] \; .
\end{eqnarray}
\noindent
In momentum space, the potential consists of the Fourier transform of the 
regularized delta-function and an infinite number of approximately local 
effective interactions (non-renormalizable) 
\begin{eqnarray}
V_{\lambda}({\bf q})&=&\left[C_0(\lambda) + C_2(\lambda) \; \frac{{\bf 
q}^2}{2\lambda^2}+C_4(\lambda) \; \frac{{\bf q}^4}{4\lambda^4}+ \dots \right] 
\; e^{-\frac{{\bf q}^2}{2\lambda^2}} \; .
\end{eqnarray}
\noindent
The effective interactions correspond to the derivatives of the delta-function 
in momentum space and model the effects of the excluded high-momentum transfer 
contributions, systematically eliminating the dependence of the observables on 
the cutoff. 

To simplify the calculations, we drop terms such as ${\bf p}.{\bf p'}$, 
obtaining an equally satisfactory effective potential that is easier to work 
with because it is completely separable:
\begin{eqnarray}
V_{\lambda}(p,p')&=&\left[C_0(\lambda) + C_2(\lambda) \; 
\frac{(p^2+p'^2)}{2\lambda^2}+C_4(\lambda) \; 
\frac{(p^4+p'^4)}{4\lambda^4}+C'_4(\lambda) \; \frac{p^2 \; p'^2}{2\lambda^4}+ 
\dots \right] \nonumber\\
&\times& e^{-\frac{p^2}{2\lambda^2}}\; e^{-\frac{p'^2}{2\lambda^2}} \; .
\label{pe}
\end{eqnarray}
\noindent
The parameters $C_i$ in the expansion can be determined order by order by 
fitting data for low-energy processes. Here we  use as ``data'' the values for 
the inverse ``on-shell'' K-matrix, calculated with an exact theory. We follow 
the method developed by Steele and Furnstahl ~\cite{steele98a}, 
determining the parameters by requiring the EFT calculation to give the same
result as the calculation using the exact theory to a  given order of
accuracy. Instead of using the T-matrix, we iterate the  effective potential
via the Lippmann-Schwinger equation for the K-matrix, which  is real and
therefore more convenient for numerical calculations of scattering 
observables. 

The Lippmann-Schwinger equation for the K-matrix with the effective potential, 
generalized to $D$ dimensions is given by
\begin{equation}
K_{\lambda}(p,p';k)=V_{\lambda}(p,p')+{\cal P}\int \; d^{D}q \; \; 
\frac{V_{\lambda}(p,q)}{k^2-q^2} \; K_{\lambda}(q,p';k) \; ,
\end{equation}
\noindent
where ${\cal P}$ denotes the principal value. In the $D=2$ case only S-wave 
scattering occurs, so the above equation refers to the S-wave potential 
(angular variable integrated out) and $d^{2}q=q \; dq$. To obtain the K-matrix 
we solve the Lippmann-Schwinger equation (non-perturbatively), using the 
technique described in Refs. \cite{phillips,richardson97,birse98a,gegelia98},
which is suitable for  separable potentials. The effective potential given in
Eq. (\ref{pe}) (truncated  at order  $p^4$) can be written in the form
\begin{equation}
V_{\lambda}(p,p')=e^{-\frac{p^2}{2\lambda^2}} \; 
e^{-\frac{p'^2}{2\lambda^2}}\sum_{i,j=0}^{2} p^{2i}\; \Lambda_{ij}\; p'^{2j} \; 
,
\end{equation}
\noindent
where $\Lambda_{ij}$ are the matrix elements of
\begin{equation}
{\bf \Lambda}=\left( \begin{array}{ccc} 
C_0(\lambda) & C_2(\lambda)/2\lambda^2  & C_4(\lambda)/4\lambda^4\\ \\
C_2(\lambda)/2\lambda^2  & C'_4(\lambda)/2\lambda^4&0\\ \\ 
C_4(\lambda)/4\lambda^4&0&0 
\end{array} \right) \; .
\end{equation}
\noindent
The solution of the Lippmann-Schwinger equation can then be written in the form

\begin{equation}
K_{\lambda}(p',p;k)=e^{-\frac{p^2}{2\lambda^2}} \; 
e^{-\frac{p'^2}{2\lambda^2}}\sum_{i,j=0}^1 p^{2i}\; \tau_{ij}(k)\; p'^{2j} \; .
\label{Kmat}
\end{equation}
\noindent
The unknown matrix ${\bf \tau}(k)$ satisfies the equation
\begin{equation}
{\bf \tau(k)}={\bf \Lambda} + {\bf \Lambda} \;{\bf {\cal I}}(k) \;{\bf 
\tau(k)}\; ,
\label{tau}
\end{equation}
\noindent
where
\begin{equation}
{\bf {\cal I}}=\left( \begin{array}{ccc} 
I_0(k) & I_1(k)  & I_2(k)\\ \\
I_1(k) & I_2(k)  & I_3(k)\\ \\ 
I_2(k) & I_3(k)  & I_4(k)
\end{array} \right) \; ,
\end{equation}

\noindent
with
\begin{equation}
I_{n}(k)=
{\cal P}\int dq \; q^{2n+D-1}\;  \frac{1}{k^2 - q^2}\; 
e^{-\frac{q^2}{\lambda^2}}\; . 
\end{equation}
\noindent
Solving Eq. (\ref{tau}) for ${\bf \tau}$ analytically  and substituting the 
solution in  Eq. (\ref{Kmat}) we obtain the effective K-matrix.

Including only the first operator (with $C_0$) in the effective potential we 
obtain
\begin{equation}
K_{\lambda}^{(0)}(p,p';k)= \frac{C_0(\lambda)}{1-C_0(\lambda) \; I_0(k)} \; 
e^{-\frac{p^2}{2\lambda^2}} \; e^{-\frac{p'^2}{2\lambda^2}}.
\end{equation} 
\noindent
Including the first two operators (with $C_0$ and $C_2$) we obtain
\begin{eqnarray}
K_{\lambda}^{(2)}(p,p';k)&=&\frac{C_0(\lambda)+\frac{C_2(\lambda)}{2\lambda^2} 
(p^2 + p'^2)+\frac{C_2^{2}(\lambda)}{2\lambda^2}\left[I_2(k)-(p^2 + p'^2)\; 
I_1(k)+p^2 \; p'^2\; I_0(k)\right]}{1-C_0(\lambda) \; I_0(k)-2 \; 
\frac{C_2(\lambda)}{2\lambda^2} \; I_1(k) - 
\frac{C_2^2(\lambda)}{2\lambda^2}[I_2(k) \; I_0(k)-I_1^2(k)]}\nonumber\\
&\times&\; e^{-\frac{p^2}{2\lambda^2}} \; e^{-\frac{p'^2}{2\lambda^2}}\; .
\end{eqnarray} 

We then fit the difference between the effective and exact inverse ``on-shell'' 
K-matrix ($p^2=p'^2=k^2$) to an interpolating polynomial in  $k^2/{\lambda^2}$ 
to highest possible order,
\begin{equation}
\Delta \; \left[\frac{1}{K}\right]=A_0+A_2 \;  \frac{k^2}{\lambda^2}+A_4 \; 
\frac{k^4}{\lambda^4}+ \cdots \; .
\label{del}
\end{equation}
\noindent
The coefficients $A_i$ are minimized with respect to the variation in the 
parameters $C_i$ of the effective potential. The number of coefficients that 
can be minimized is given by the number of parameters appearing in the 
effective potential. This method is robust and numerically more stable than the 
matching at discrete momenta used by Lepage, allowing extension to higher 
orders ~\cite{steele98a}. 

By adjusting only the parameter $C_0(\lambda)$ we should eliminate the leading 
error. As each term is added to the effective potential, followed by the 
adjustment of the respective parameter, we expect the errors in the 
``on-shell'' K-matrix to be systematically reduced by {\it powers of 
$k^2/{\lambda^2}$}. Thus, in a log-log plot for $\Delta[1/K]$ we expect to 
obtain straight lines with slope given by the dominant power of $k/{\lambda}$ 
in the error. We know of no rigorous proof of this expectation in EFT,
although it may  be possible to show that each $A_i$ can be taken to zero with
a complete set of  operators to the appropriate order in $p^2/\lambda^2$ in the
hamiltonian. Note,  however, that the number of couplings, $C_i$, grows more
rapidly than the  number of $A_i's$ and the relation between them is
nonlinear. We assume that we  can adjust operators of order
$(p^2/\lambda^2)^m$ to remove errors in the  inverse ``on-shell'' K-matrix of
order $(k^2/\lambda^2)^m$, and all of our  results confirm this.

The expansion given by Eq.( \ref{del}), at best, becomes invalid when the 
momenta involved are of the same order as the cutoff, where one expects the 
short distance effects to be directly resolved. This point corresponds to the 
radius of convergence of the effective theory. 

The identification of this power counting for the order of errors implies that 
the same systematic dependence on momentum is expected in the evaluation of 
errors for other low-energy observables. It is important to observe that this 
power counting applies to the predictions for the observables and not to the 
potential. As new terms are added and the corresponding couplings are adjusted, 
the lower-order couplings change. This occurs because the truncated potential 
still contains all orders in $(p/{\lambda})^2$ due to the regulating function, 
so when the Lippmann-Schwinger equation is solved non-perturbatively all powers 
of momentum are generated and there is no simple linear relationship between 
powers of momentum in the potential and powers in the scattering observables.

One of our goals is to compare the EFT and the SRG methods. In principle, this 
could be achieved by comparing the results of the error analysis for low-energy 
scattering observables obtained in each  case. However, the calculation of the 
K-matrix (or T-matrix) with the renormalized potential obtained using Wegner's 
transformation \cite{wegner} requires the numerical solution of an integral 
equation involving a non-separable kernel, which is unstable when simple 
techniques are applied. Since we expect the bound-state energy  and the 
low-energy scattering observables to have similar error scaling (after fixing 
the effective potential to a given order), we focus on the bound state problem, 
which is numerically much more stable and allows a satisfactory error analysis. 
Using the effective potential (truncated at a given order) with the parameters 
fixed by fitting the inverse ``on-shell'' K-matrix, we diagonalize the 
effective hamiltonian numerically, obtaining the bound-state energies. This is 
implemented by discretizing the Schr\"{o}dinger equation.
%

%

%

\section{Similarity Renormalization Group}

In this section we review the general formulation of the SRG developed by 
G{\l}azek and Wilson ~\cite{wilgla1,wilgla2} and a specific transformation
developed by  Wegner ~\cite{wegner}. The reader may wish to skip the general
formulation on a  first reading.

\subsection{G{\l}azek-Wilson Formulation}

Consider a system described by a hamiltonian written in the form
\begin{equation}
H=h+V \; ,
\end{equation}
where $h$ is the free hamiltonian and $V$ is an interaction.

In general, the hamiltonian can couple states of all energy scales and such  
couplings can be a source of ultraviolet divergences. The goal of the SRG is to 
obtain an effective hamiltonian in which the couplings  between high and 
low-energy states are removed, while avoiding any problems from small energy 
denominators in effective interactions. The procedure is implemented by a
unitary transformation that  generates effective interactions that reproduce
the effects of the eliminated  couplings. The effective hamiltonian  cannot
produce ultraviolet divergences at  any order in perturbation theory as long
as its matrix elements are finite. 

In our discussion we will use the basis of eigenstates of the free hamiltonian,

\begin{equation}
h |i >=\epsilon_i|i > \; .
\end{equation}
\noindent
We start by defining a bare hamiltonian, $H_\Lambda$, regulated by a very large 
cutoff $\Lambda$ (here with dimensions of energy) on the change in free energy 
at the interaction vertices,
\begin{eqnarray}
H_\Lambda&\equiv&h+V_{\Lambda} \;,\\
V_{\Lambda}&\equiv& f_{\Lambda} {\overline V}_{\Lambda}\;,\\
{\overline V}_{\Lambda}&\equiv& v+H_{\Lambda}^{ct}
\;,
\end{eqnarray}
where $f_{\Lambda}$ is a ``similarity function", ${\overline V}_{\Lambda}$ is 
defined as the reduced interaction and $H_{\Lambda}^{ct}$ are counterterms that 
must be determined through the process of renormalization in order to remove 
$\Lambda$ dependence in physical quantities.   

The similarity function $f_{\Lambda}$ regulates the hamiltonian by suppressing 
matrix elements between free states with significantly large energy difference 
and acts in the following way:
\begin{eqnarray}
< i | f_{_{\Lambda}} H_{_{\Lambda}} |j > &\equiv& \epsilon_i \; 
\delta_{ij}+f_{_{\Lambda }}(\epsilon_i-\epsilon_j) < i |{\overline 
V}_{_{\Lambda}} |j > \nonumber\\
&\equiv& \epsilon_i \; \delta_{ij}+ f_{_{\Lambda ij}} {\overline V}_{_{\Lambda 
ij}}.
\end{eqnarray}
\noindent
\noindent
Typically, the similarity function is chosen to be a smooth function 
satisfying 
\begin{eqnarray}
&&(i) f_{_{\Lambda }}(\epsilon_i-\epsilon_j) \rightarrow 1, \; {\rm when} \; 
|\epsilon_i-\epsilon_j|<< \Lambda \; ,\nonumber\\
&&(ii)f_{_{\Lambda }}(\epsilon_i-\epsilon_j)  \rightarrow 0, \; {\rm when} \; 
|\epsilon_i-\epsilon_j| >>  \Lambda \; .
\end{eqnarray}
\noindent
In several papers the similarity function has been chosen to be a step 
function.  Although  useful for doing analytic calculations, such a choice can 
lead to pathologies ~\cite{billy}.

The similarity transformation  is defined to act on the bare regulated 
Hamiltonian, $H_\Lambda$, lowering the cutoff down to a scale $\lambda$:
\begin{eqnarray}
H_\lambda&\equiv& U(\lambda ,\Lambda)\; H_\Lambda \; U^\dagger(\lambda, 
\Lambda)\;.
\label{eq:star}
\end{eqnarray}
\noindent
The renormalized Hamiltonian can be written in the general form
\begin{eqnarray}
H_{\lambda} &\equiv &h+V_{\lambda} \;, \\
V_{\lambda} &\equiv& f_{\lambda}\; {\overline V}_{\lambda}\;.
\end{eqnarray}
\noindent
The transformation is unitary, so  $H_\Lambda$ and $H_\lambda$ produce the same 
spectra for observables. Also, if an exact transformation is implemented, the 
physical predictions using the renormalized Hamiltonian must be independent of 
the cutoff $\lambda$ and $ H_{\Lambda}^{ct}$ is chosen so that they also 
become independent of $\Lambda$ as $\Lambda \rightarrow \infty$.

The unitarity condition is given by:
\begin{eqnarray}
U(\lambda, \Lambda)\;  U^\dagger(\lambda, \Lambda)&\equiv&U^\dagger(\lambda, 
\Lambda)\; U(\lambda, \Lambda)  \equiv 1
\;.\label{eq:unitary}
\end{eqnarray}
\noindent
The similarity transformation $U$ can be defined in terms of an anti-hermitian 
operator $T_{\lambda}$ ($T_\lambda^\dagger=-T_\lambda$) which generates 
infinitesimal changes of the cutoff energy scale,
\begin{eqnarray}
U(\lambda ,\Lambda)&\equiv&{\cal T} \exp \left(\int_\lambda^\Lambda 
T_{\lambda^{\prime}}\; d \lambda^{\prime}\right)\; ,
\label{eq:tdef}
\end{eqnarray}
where ${\cal T}$ orders operators from left to right in order of {\it 
increasing} energy scale $\lambda^{\prime}$. Using
\begin{equation} 
T_\lambda=U(\lambda ,\Lambda)\; \frac{dU^\dagger(\lambda ,\Lambda)}{d \lambda}
  =-\frac{dU(\lambda ,\Lambda)}{d \lambda}\; U^\dagger(\lambda ,\Lambda)\;,
\end{equation}
\noindent
and the unitarity condition Eq. (\ref{eq:unitary}), we can write 
Eq. (\ref{eq:star})  in a differential form,
\begin{eqnarray}
\frac{d H_{\lambda}}{d \lambda}&=& \left [ H_\lambda,T_\lambda\right] 
\label{eq:star2}\; .
\end{eqnarray}

This is a first-order differential equation, which is solved with the boundary 
condition  $H_\lambda |_{_{\lambda \rightarrow \Lambda}} \equiv H_\Lambda$.
The  bare Hamiltonian is typically given by the canonical Hamiltonian plus 
counterterms that must be uniquely fixed to complete the renormalization.  

The operator $T_\lambda$ is defined by specifying how 
${\overline V}_\lambda$ and $h$ depend on the  cutoff scale $\lambda$. For 
simplicity in this paper, we demand that $h$ is independent of $\lambda$, 
although this may not lead to an increasingly diagonal effective hamiltonian in 
all cases. We also demand that no small energy denominators can appear in the 
hamiltonian. These constraints are implemented by the conditions
\begin{eqnarray}
\frac{d h}{d \lambda}&\equiv&0\;,
\label{eq:star4}\\ \nonumber\\
\frac{d {\overline V}_\lambda}{d \lambda}&\equiv& [V_\lambda,T_\lambda]\;.
\label{eq:star3}
\end{eqnarray}

To obtain the renormalized Hamiltonian perturbatively, expand
\begin{eqnarray}
&&{\overline V}_{\lambda}=
{\overline V}_{\lambda}^{^{(1)}}
+{\overline V}_{\lambda}^{^{(2)}}+\cdots
\label{eq:yes}~,
\\
&&T_{\lambda}=
T_{\lambda}^{^{(1)}}
+T_{\lambda}^{^{(2)}}+\cdots~,\\
&&H_{_{\Lambda}}^{^{ct}} = H_{_{\Lambda}}^{^{(2),ct}}
+H_{_{\Lambda}}^{^{(3),ct}} +\cdots\;,
\end{eqnarray}
where the superscripts denote the order in the original interaction, $V$. A 
general form of these effective interactions is
\begin{eqnarray}
{\overline V}_\lambda^{^{(i)}}&=&-\sum_{j,k=1}^\infty \delta_{(j+k,i)}
\int_\lambda^\Lambda d\lambda^\prime \; [ V_{\lambda^\prime}^{^{(j)}} , 
T_{\lambda^\prime}^{^{(k)}}]
+ H_{_{\Lambda}}^{^{(i),ct}}
 \;,
\end{eqnarray}
 for $i=2,3, \cdots$, with ${\overline V}_\lambda^{^{(1)}}=v$. For instance, 
the explicit form of the second-order effective
interaction ${\overline V}_\lambda^{^{(2)}}$ is
\begin{equation}
{\overline V}_{\lambda ij}^{^{(2)}}=
\sum_{k} V_{ik} 
{V}_{kj} \left(\frac{g^{(\lambda \Lambda)}_{ikj}}{\Delta_{ik}} +
  \frac{g^{(\lambda \Lambda)}_{jki}}{\Delta_{jk}}
  \right) +H_{\Lambda ij}^{^{(2),ct}}\;,
\end{equation}
\noindent
where
\begin{eqnarray}
&&g^{(\lambda \Lambda)}_{ikj} \equiv
  \int_\lambda^\Lambda \; d \lambda^\prime \; f_{\lambda^\prime jk} 
  \; \frac{d f_{\lambda^\prime ki}}{d \lambda^\prime}\; ,
\\
&&\Delta_{ij}=\epsilon_i-\epsilon_j \; .
\end{eqnarray}

The counterterms $H_{\Lambda}^{^{(n),ct}}$ can be determined order-by-order 
using the idea of coupling-coherence ~\cite{ochme,coupcoh}. This is 
implemented by requiring the hamiltonian to reproduce itself in form under the 
similarity transformation, the only change being explicit dependence on the 
running cutoff in the operators and the implicit cutoff dependence in a finite 
number of independent running couplings. All other couplings depend on the 
cutoff only through their dependence on the independent couplings.  In general,  
we also demand  the dependent couplings to vanish when the independent 
couplings are taken to zero; i.e, the interactions are turned off. If the only 
independent coupling  in the theory is $\alpha_{\lambda}$, the renormalized 
hamiltonian can be written as an expansion in powers of this coupling:
\begin{equation}
H_{\lambda}=h+\alpha_{\lambda}{\cal O}^{(1)}+\alpha_{\lambda}^2{\cal O}^{(2)}+ 
\dots \; .
\end{equation}
\noindent
In this way, the effective hamiltonian obtained using the similarity 
transformation is completely determined by the underlying theory. The procedure 
can be extended to arbitrarily high orders, although it becomes increasingly 
complex both analytically and numerically. 

%


\subsection{Wegner Formulation}

The Wegner formulation of the SRG \cite{wegner} is defined in a very elegant 
way in terms of a flow equation analogous to the SRG Equation in the 
G{\l}azek-Wilson formalism \cite{wilgla1,wilgla2},
\begin{equation}
\frac{d H_s}{ds}=[H_s,T_s]
\;.
\end{equation}
\noindent
Here the hamiltonian $H_s=h+v_s$ evolves with a flow parameter $s$ that ranges 
from $0$ to $\infty$. The flow-parameter has dimensions $1/({\rm energy})^2$ 
and is given in terms of the similarity cutoff $\lambda$ by $s=1/\lambda^2$. 

In Wegner's  scheme the similarity transformation in defined by an explicit 
form for the generator of the similarity transformation, $T_s=[H_s,H_0]$, which  
corresponds to the choice of a gaussian similarity function with uniform width. 
In the original formulation,
Wegner advocates the inclusion of the full diagonal part of the hamiltonian at 
scale $s$ in  $H_0$. For a perturbative calculation of $H_s$, we can use
the free  hamiltonian, $H_0=h$. With this choice, the flow equation for the
hamiltonian is  given by 
\begin{equation}
\frac{d H_s}{ds}=[H_s,[H_s,h]]\; .
\end{equation}

The reduced interaction, ${\overline V}_{sij}$ (the interaction with the 
gaussian similarity function factored out) is defined by
\begin{eqnarray}
V_{sij}= f_{_{sij}}\; {\overline V}_{sij}
\;, \\
f_{_{sij}}=e^{-s\Delta_{ij}^2} \; .
\end{eqnarray}
Assuming that the free hamiltonian is independent of $s$, we obtain the flow 
equation for the reduced interaction,
\begin{equation}
\frac{d{\overline V}_{sij}}{ds}=\sum_k\left(\Delta_{ik}+\Delta_{jk}\right)
\; {\overline V}_{sik} \; {\overline V}_{skj} \; e^{-2s
\Delta_{ik}\Delta_{jk}}
\;,
\label{eq:wegnerbigone}
\end{equation}
\noindent
where we use 
$\Delta_{ij}^2-\Delta_{ik}^2-\Delta_{jk}^2=-2\Delta_{ik}\Delta_{jk}$.
We should emphasize that this is an exact equation.

To solve this equation we impose a 
boundary condition, $H_s |_{_{s \rightarrow s_0}} \equiv H_{s_0}$. 
Then, we make a perturbative expansion,
\begin{equation}
{\overline V}_{s}=
{\overline V}_{s}^{^{(1)}}
+{\overline V}_{s}^{^{(2)}}+\cdots\;,
\end{equation}
where the superscript implies the order in the bare interaction ${\overline 
V}_{s_{_{0}}}$. It is important to observe that counterterms are implicit in 
the bare interaction and can be determined in the renormalization process using 
coupling coherence.

At first order we have
\begin{equation}
\frac{d{\overline V}_{s_{ij}}^{^{(1)}}}{ds}=0\;,
\end{equation} 
which implies
\begin{equation}
{\overline V}_{sij}^{^{(1)}}={\overline V}_{s_{_{0}}ij}
\;,
\end{equation}
where $s$ is the final scale. Because of the dimensions of the flow parameter 
we have $s > s_{_{0}}$, corresponding to a smaller cutoff.  
The ``no cutoff limit" corresponds to $s_{_{0}}\longrightarrow 0$.

At second order we have
\begin{equation}
\frac{d{\overline 
V}_{sij}^{^{(2)}}}{ds}=\sum_k\left(\Delta_{ik}+\Delta_{jk}\right)
\; {\overline V}_{s_{_{0}}ik}\; {\overline V}_{s_{_{0}}kj} \; e^{-2s
\Delta_{ik}\Delta_{jk}}
\;.
\end{equation}
Integrating, we obtain
\begin{eqnarray}
{\overline V}_{sij}^{^{(2)}}&=&\frac{1}{2}\sum_k
{\overline V}_{s_{_{0}}ik}\; {\overline V}_{s_{_{0}}kj} \; 
\left(\frac{1}{\Delta_{ik}}+\frac{1}{\Delta_{jk}}\right)\times\nonumber\\
&&~~~~~~~~~~\times \left[e^{-2 s_0 
\Delta_{ik}\Delta_{jk}}-e^{-2s\Delta_{ik}\Delta_{jk}}\right]
\;.
\end{eqnarray}

By construction, the Wegner transformation is unitary and avoids small energy 
denominators. The Wegner transformation is one of the G{\l}azek-Wilson 
transformations, with the similarity function chosen to be $f_{_{\lambda 
ij}}=e^{-\Delta_{ij}^2/\lambda^2}$. 
%

%

\subsection{Strategy}

In our applications of the SRG we use Wegner's transformation. The renormalized 
hamiltonian for the non-relativistic delta-function potential in D-dimensions 
is given by 
\begin{equation}
H_{\lambda}({\bf p},{\bf p'})=p^2 \delta^{(D)}({\bf p}-{\bf p'}) + 
e^{-\frac{(p^2-p'^2)^2}{\lambda^4}}\; \left[{\bar V}_{\lambda}^{(1)}({\bf 
p},{\bf p'})+{\bar V}_{\lambda}^{(2)}({\bf p},{\bf p'})+... \right]\; ,
\label{renh}
\end{equation}
\noindent
where
\begin{eqnarray}
{\bar V}_{\lambda}^{(1)}({\bf p},{\bf 
p'})&=&-\frac{\alpha_{\lambda,i}}{(2\pi)^D} \; ,\\
{\bar V}_{\lambda}^{(2)}({\bf p},{\bf p'})&=&\alpha_{\lambda,i}^2 \; 
F^{(2)}_{s}({\bf p},{\bf p'}) \; , \\
{\bar V}_{\lambda}^{(n)}({\bf p},{\bf p'})&=&\alpha_{\lambda,i}^n \; 
F^{(n)}_{s}({\bf p},{\bf p'}) \; .
\end{eqnarray}
\noindent
Here $\lambda$ is a momentum cutoff (as opposed to the energy cutoff
discussed above) related to the flow parameter by 
$s=1/{\lambda^4}$. The index $i$ denotes the order of the calculation for the 
running coupling.

The renormalized hamiltonian can be used to compute eigenvalues and 
eigenstates. Since the hamiltonian is derived perturbatively we expect cutoff 
dependent errors in the observables. Formally, we can regroup the terms in the 
renormalized hamiltonian and write it as a momentum  expansion similar to the 
one for EFT in Eq.~(\ref{pe}). The difference is that the expansion 
parameters are analytic functions of the running coupling $\alpha_{\lambda}$. 
Expanding  the operators $F^{(n)}_{s}({\bf p},{\bf p'})$ in powers of 
$p^2/{\lambda^2}$ we obtain
\begin{equation}
F^{(n)}_{s}({\bf p},{\bf p'})=z_0+z_2 \; \frac{(p^2+p'^2)}{2\lambda^2}+z_4 \; 
\frac{(p^4+ p'^2)}{4\lambda^4}+ z_4' \; \frac{p^2 p'^2}{2\lambda^4}+\cdots \; ,
\end{equation}
where the ${z_i}^{'}s$ are constants. Regrouping the terms we obtain
\begin{equation}
H_{\lambda}(p,p')=p^2 \delta^{(D)}({\bf p}-{\bf p'}) + 
e^{-\frac{(p^2-p'^2)^2}{\lambda^4}}\; 
\left[g_0(\alpha_{\lambda})+g_2(\alpha_{\lambda})\frac{( 
p^2+p'^2)}{2\lambda^2}+ \dots  \right]\; ,
\label{mexp}
\end{equation}
\noindent
where
\begin{equation}
g_i(\alpha_{\lambda})=a_i(\lambda)\; \alpha_{\lambda}+b_i(\lambda)\; 
\alpha_{\lambda}^2 + \dots \; .
\label{coupexp}
\end{equation}
\noindent
For $D=2$ all $a_i(\lambda)$, $b_i(\lambda)$, etc. are simply constants, while 
for $D=1$ they are functions of $\lambda$ because of the implicit mass scale 
which is hidden when we set $2m=1$.

We can identify three interdependent sources of errors in the perturbative 
similarity renormalization group when the hamiltonian given by Eq.~(\ref{mexp}) 
is truncated and used to compute a physical quantity:
\vspace{0.5cm}

\noindent
a) errors introduced by the truncation of the hamiltonian at a given order in 
$p^2/\lambda^2$, which are analogous to the errors in EFT;
\vspace{0.3cm}

\noindent
b) errors introduced by the truncation of the hamiltonian at a given order in 
the running coupling $\alpha_{\lambda,i}$, which correspond to the use of an 
approximation for the functions $g_i$;
\vspace{0.3cm}

\noindent
c) errors introduced by the approximation for the running coupling 
$\alpha_{\lambda,i}$.
\vspace{0.5cm}

In the actual calculation using the hamiltonian given by Eq.~(\ref{renh}) 
errors of type (a) do not appear directly because we do not truncate the
operators that appear in the hamiltonian. However, errors of type (b) can be
understood as coming from approximating the couplings in front of the
operators in Eq.~(\ref{mexp}). In EFT these couplings are fit to data, but we
approximate them as in Eq.~(\ref{coupexp}) and this leads to an approximate
cancellation of the power-law errors completely removed by these operators in
EFT.  Errors of type (c) appear in our calculations only because we do not
fit the canonical coupling to data at each scale, but fix it at a given
scale and evolve it perturbatively from that scale. The strategy  we would use
for a realistic   theory ({\it e.g.}, QED and QCD) is the following:
\vspace{0.5cm}

\noindent
1) Obtain the renormalized hamiltonian using the similarity transformation and 
coupling-coherence, truncating the hamiltonian at a given order in powers of
$\alpha_{\lambda,i}$.
\vspace{0.3cm}

\noindent
2) Fix the coupling $\alpha_{\lambda}$ by fitting an observable ({\it e.g.}, a 
bound-state energy).
\vspace{0.3cm}

\noindent
3) Evaluate other observables ({\it e.g.}, scattering phase shifts). 
\vspace{0.5cm}

As pointed out before, the evaluation of scattering observables with the 
similarity hamiltonian with standard techniques is complicated and so in our 
examples  we focus on the bound state errors. We fix the coupling at some scale  
using a given renormalization  prescription  and use the flow-equation to 
obtain the coupling as a function of the cutoff $\lambda$ to a given order. We 
then perform a sequence of bound-state calculations with better approximations 
for the hamiltonian such that the errors in the bound-state energy are 
systematically reduced. Once the sources of errors are identified, it becomes 
relatively simple to analyze order-by-order how such errors scale with the 
cutoff $\lambda$. In principle, to completely eliminate the errors proportional 
to some power $m$ in the momentum expansion we should use the similarity 
hamiltonian with the exact running coupling (renormalized to all orders) and 
include the contributions up to ${\cal O}(p^m/{\lambda^m})$ coming from all 
effective interactions (all orders in $\alpha_{\lambda}$). Since we use a 
perturbative approximation for the running coupling and only a finite number of 
effective interactions, we do not expect the scaling analysis for the errors in 
SRG to be as simple as in the EFT approach. Some details of this scaling 
analysis are presented later for the specific examples we work out. We should 
emphasize that in a realistic calculation we would fit the coupling 
$\alpha_{\lambda}$ to an observable. This nonperturbative renormalization 
eliminates the dominant source of errors we display in SRG calculations in this 
paper. We choose to renormalize the coupling perturbatively in this paper 
because the only observable we compute is the single bound state energy of a 
delta-function potential, and fitting this energy would prevent us from 
displaying errors. 

%

%

\section{One-Dimensional Delta-function Potential}

Consider a system of two non-relativistic particles in one dimension 
interacting via an attractive Dirac delta-function, contact or zero range 
potential.  The Schr\"{o}dinger equation in position space (with $2 m=1$ and
$\hbar=1$),  separating out the center of mass motion, is given by
\begin{equation}
-\frac{d^2 \Psi(x)}{dx^2}-\alpha_0 \;\delta^{(1)}(x) \; \Psi(x)=E \; \Psi(x)\; 
.
\end{equation}
\noindent
The coupling constant $\alpha_0$ completely fixes the underlying theory.
%


%

\subsection{Exact Solution}

The exact solutions of this equation are familiar to all students of quantum 
mechanics. There is one bound state, $(E < 0)$  with binding energy 
$E_0=-E=\frac{\alpha_0^2}{4}$ and normalized wave-function
\begin{equation}
\Psi(x)=\sqrt{\frac{\alpha_0}{2}} \; e^{-\frac{\alpha_0}{2}\; |x|} \; .
\end{equation}
\noindent
In the scattering problem $(E=k^2 > 0)$ the Schr\"{o}dinger equation is 
equivalent to a Lippmann-Schwinger equation, whose solution with outgoing waves 
is given by
\begin{equation}
\Psi_{k}(x)=\frac{e^{i k x}}{\sqrt{2 \pi}} \; \left[\frac{2 i k}{2 i k + 
\alpha_0}\right] \; .
\end{equation}

The problem is easily solved in momentum space, also. We do this to lay the 
groundwork for our first illustration of renormalization techniques. The 
corresponding Schr\"{o}dinger equation in momentum space is given by
\begin{equation}
p^2 \;  \Phi(p)-\frac{\alpha_0}{2\pi}\; \int_{-\infty}^{\infty} \; dq 
\;\;\Phi(q)=E \; \Phi(p) \; ,
\label{semom1}
\end{equation}
\noindent
where $\Phi(p)$ is the Fourier transform of the position space wave-function,
\begin{equation}
\Phi(p)=\frac{1}{\sqrt{2\pi}}\int_{-\infty}^{\infty} \; dx  \;\; \Psi(x)\;e^{-i 
p x} \; .
\end{equation}
\noindent
Identifying the integral in Eq. (\ref{semom1}) as the position space 
wave-function at the origin
\begin{equation}
\frac{1}{\sqrt{2\pi}}\int_{-\infty}^{\infty} \; dq \;\; \Phi(q)=\Psi(0) \; ,
\end{equation}
\noindent
we obtain
\begin{equation}
p^2 \;  \Phi(p)-\frac{\alpha_0}{\sqrt{2\pi}}\; \Psi(0)=E \; \Phi(p) \; .
\end{equation}

First, we consider the bound-state problem.  By rearranging the terms, we 
obtain
\begin{equation}
\Phi(p)=\frac{\alpha_0}{\sqrt{2\pi}}\; \frac{\Psi(0)}{(p^2+E_0)} \; .
\end{equation}
\noindent
Integrating the above equation in $p$, we obtain
\begin{equation}
1=\frac{\alpha_0}{2\pi}\; \int_{-\infty}^{\infty} \; dp \;\;\frac{1}{(p^2+E_0)} 
\; .
\label{bsi}
\end{equation}
\noindent
This integral fixes the binding energy, $E_0=\frac{\alpha_0^2}{4}$. The 
corresponding bound-state wave-function in momentum space is 
\begin{equation}
\Phi(p)=\frac{1}{\sqrt{2\pi}}\; \frac{2 \;E_{0}^{\frac{3}{4}}}{(p^2+E_0)} \; .
\end{equation}
\noindent
Solving for the scattering states we obtain the solution corresponding to 
outgoing waves 
\begin{equation}
\Phi_{k}(p)=\delta(p-k)+\frac{\alpha_0}{2 \pi}\; \left[\frac{2 i k}{2 i k + 
\alpha_0}\right]\frac{1}{(p^2-k^2-i\epsilon)} \; ,
\end{equation}
\noindent
where $k=\sqrt{E}$.

All scattering observables can be obtained from the T-matrix, which is given by 
a  Lippmann-Schwinger equation:
\begin{equation}
T(p,p';k)=V(p,p')+\int_{-\infty}^{\infty} \; dq \; \; \frac{V(p,q)}{k^2-q^2+i 
\epsilon} \; T(q,p';k) \; .
\end{equation}
\noindent
The ``on-shell'' $(p^2=p'^2=k^2)$ T-matrix for a Dirac delta-function potential 
is
\begin{eqnarray}
T_0(k)=-\frac{\alpha_0}{2\pi}\left[\frac{2 i k}{2ik+\alpha_0}\right] \; .
\end{eqnarray}

A more convenient way to treat the scattering problem for our purposes is to 
use the K-matrix, which satisfies the Lippmann-Schwinger equation 
\begin{equation}
K(p,p';k)=V(p,p')+{\cal P}\int_{-\infty}^{\infty} \; dq \; \; 
\frac{V(p,q)}{k^2-q^2} \; K(q,p';k) \; .
\end{equation}
\noindent
This equation is  similar to that for the T-matrix, except standing-wave 
boundary conditions are imposed so that the usual $i\epsilon$ prescription is 
replaced by the principal value ${\cal P}$. As a result, the K-matrix is  real. 
The on-shell K-matrix is given by
\begin{equation}
K_0(k)=-\frac{\alpha_0}{2 \pi} \; .
\end{equation}
\noindent
The ``on-shell'' K-matrix and T-matrix  are related by the expression
\begin{equation}
K_0(k)=\frac{T_0(k)}{1-\frac{i\pi}{k}T_0(k)} \; .
\end{equation}
\noindent
Using either
\begin{equation}
k \; {\rm cot}\;\delta_{0}(k)-i k=-\frac{k^2}{\pi}\frac{1}{T_0(k)} \; ,
\end{equation}
\noindent
or
\begin{equation}
k \; {\rm cot}\; \delta_{0}(k)=-\frac{k^2}{\pi}\frac{1}{K_0(k)}\; ,
\end{equation}
\noindent
the phase-shifts are easily calculated:
\begin{equation}
{\rm cot}\; \delta_{0}(k)=\frac{2 k}{\alpha_0} \; .
\end{equation}

Although the one-dimensional problem can be solved exactly without
regularization and renormalization, for the sake of further comparison
with the two-dimensional case and to illustrate the SRG and EFT techniques, it
is worth analyzing what happens when a momentum cutoff is introduced. If the
integral in Eq. (\ref{bsi}) is regulated by a cutoff $\Lambda$,
\begin{equation}
1=\frac{\alpha_0}{2\pi}\; \int_{-\Lambda}^{\Lambda} \; dp 
\;\;\frac{1}{(p^2+E_0)} \; ,
\end{equation}
\noindent
we obtain a transcendental equation, whose solution gives a cutoff-dependent
binding energy:
\begin{equation}
\frac{\pi}{\alpha_0}\; \sqrt{E_0(\Lambda)}= {\rm 
arctg}\left(\frac{\Lambda}{\sqrt{E_0(\Lambda)}}\right) \; .
\end{equation}
\noindent
Clearly, the cutoff dependence of the bound state energy is  eliminated if we 
let the coupling run like 
\begin{equation}
\alpha_0 \rightarrow \alpha_{\Lambda}=\frac{\pi \; \sqrt{E_0}}{{\rm arctg} \; 
\left(\frac{\Lambda}{\sqrt{E_0}}\right)} \; .
\end{equation}
\noindent
For $\Lambda >> \sqrt{E_0}$ this cutoff dependence can be eliminated 
perturbatively by using the expansion of $\alpha_{\Lambda}$ in powers of 
$\sqrt{E_0}/{\Lambda}$ to a given order: 
\begin{equation}
\alpha_{\Lambda}=2 \sqrt{E_0}\; 
\left[1+\frac{2}{\pi}\;\frac{\sqrt{E_0}}{\Lambda}+\frac{4}{\pi^2}\;
\frac{E_0}{\Lambda^2}+
{\cal O}\left(\frac{E_0^{3/2}}{\Lambda^3}\right)\right] \; .
\end{equation}
\noindent
In terms of the ``bare'' coupling, $\alpha_0$, we obtain
\begin{equation}
\alpha_{\Lambda}=\alpha_0\; \left[1+\frac{1}{\pi}\;\frac{\alpha_0}{\Lambda}+
\frac{1}{\pi^2}\;\frac{\alpha_0^2}{\Lambda^2}+
{\cal O}\left(\frac{\alpha_0^3}{\Lambda^3}\right)\right] \; .
\end{equation}
\noindent
Of course $\alpha_{\Lambda} \rightarrow \alpha_0$ when $\Lambda \rightarrow 
\infty$.

%

%

\subsection{Effective Field Theory Method}

We now apply the EFT approach to the one-dimensional delta-function problem. 
The power counting scheme in this  case is very simple and can be 
understood easily by analyzing how the errors scale with the cutoff $\lambda$.
For  simplicity, we consider the errors in the inverse K-matrix. 

As described before, the EFT method is designed to systematically eliminate 
errors order by order in $k/{\lambda}$. Beginning with the leading-order 
prediction, we use the effective potential with one parameter. Evaluating the 
difference
$\Delta \; [1/K^{(0)}]=1/K_{\lambda}^{(0)}-1/K_0$ and expanding it in powers of 
$k/{\lambda}$ we obtain:
\begin{equation}
\Delta \; \left[\frac{1}{K^{(0)}}\right]=
\left(\frac{1}{C_0}-\frac{2\; \sqrt{\pi}}{\lambda}+\frac{2\pi}{\alpha_0}\right)
+\left(\frac{1}{C_0}-\frac{2 \; \sqrt{\pi}}{3 \; \lambda}\right)\; 
\frac{k^2}{\lambda^2}+{\cal O}\left(\frac{k^4}{\lambda^4}\right)\; .
\end{equation}
\noindent
Note that the regulator generates only power-law errors, which indicates that 
there are no divergences as $\lambda \rightarrow \infty$ and renormalization is
required only to remove finite  errors. Clearly, the leading order error is
eliminated by choosing
\begin{equation}
C_0=\frac{-\alpha_0/{2\pi}}{1-\frac{1}{\sqrt{\pi}}\frac{\alpha_0}{\lambda}} \; 
,
\end{equation}
\noindent
and the remaining errors are dominated by the term 
\begin{equation}
{\cal O}\left(\frac{k^2}{\lambda^2}\right)
=\left(-\frac{2\pi}{\alpha_0}+\frac{4\; 
\sqrt{\pi}}{3\lambda}\right)\frac{k^2}{\lambda^2}\; .
\end{equation}

For the next-to-leading order prediction, we use the effective potential with 
two parameters, obtaining the expansion
\begin{equation}
\Delta \; \left[\frac{1}{K^{(2)}}\right]=A_0+A_2 \; \frac{k^2}{\lambda^2}+{\cal 
O}\left(\frac{k^4}{\lambda^4}\right)\; ,
\end{equation}
\noindent
where
\begin{eqnarray}
A_0&=&\frac{-2\left(1+\frac{C_2\; \sqrt{\pi}}{2 \; 
\lambda}\right)^2+2\left(\frac{2 \; \sqrt{\pi} \; C_0}{\lambda}-\frac{C_2^2 \; 
\pi}{4\;\lambda^2}\right)}{-2\; C_0+\frac{C_2^2 \; 
\sqrt{\pi}}{4\;\lambda}}+\frac{2\pi}{\alpha_0}\\ \nonumber\\ \nonumber\\
A_2&=& -\frac{8\; C_0^2 \; \sqrt{\pi}+4\;C_0\; \lambda \left(3+3 \; \frac{C_2\; 
\sqrt{\pi}}{\lambda}+\frac{5 \;C_2^2 \; \pi}{4\; \lambda^2}\right)}{3\; 
\lambda\; \left(-2 \; C_0+\frac{C_2^2 \; \sqrt{\pi}}{4\; 
\lambda}\right)^2}\nonumber\\ \nonumber\\
&-&\frac{C_2\; \lambda \; \left(12+\frac{33 \; C_2\; \sqrt{\pi}}{2 \; 
\lambda}+\frac{15 \; C_2^2 \; \pi}{2 \; \lambda^2}+\frac{5 \; C_2^3 \; 
\pi^{3/2}}{4 \; \lambda^3}\right)}{3\; \lambda\; \left(-2 \; C_0+\frac{C_2^2 
\;\; \sqrt{\pi}}{4\; \lambda}\right)^2} \; .
\end{eqnarray}

The parameters $C_0$ and $C_2$ are determined by solving the system of two 
coupled non-linear equations given by
\begin{equation}
A_0=0 \; , \; \; A_2=0 \; .
\end{equation}
\noindent
One of the solutions gives a parameter $C_0$ that diverges when $\lambda 
\rightarrow \infty$ and so is not acceptable. The other solution, 
\begin{eqnarray}
C_0&=&\frac{-27 \sqrt{\pi}\; \alpha_0 \; \lambda^2
+16 \; \alpha_0^2 \; \lambda
+12 \; \lambda^3 \; \pi}{2\sqrt{\pi}(4 \; \alpha_0^2-9 \; \alpha_0 \; \lambda 
\; \sqrt{\pi}+6 \; \lambda^2 \; \pi)}\nonumber\\ \nonumber\\
&-&\frac{\sqrt{
\frac{6 \; \lambda^2}{\pi}\; \left(\alpha_0
-\lambda \; \sqrt{\pi}\right)^2 \; \left(4 \; \alpha_0^2
-9 \; \alpha_0 \; \lambda \; \sqrt{\pi}+6 \; \lambda^2 \; \sqrt{\pi} \right)}}
{4 \; \alpha_0^2-9 \; \alpha_0 \; \lambda \; \sqrt{\pi}+6 \; \lambda^2 \; \pi} 
\; , \\   \nonumber\\ \nonumber\\
C_2&=&\frac{-8\;\alpha_0^2 \; \sqrt{\pi}\; \lambda + 18 \; \alpha_0 \; \pi \; 
\lambda^2-12 \; \lambda^3 \; \pi^{3/2}}{\pi(4 \; \alpha_0^2-9 \; \alpha_0 \; 
\lambda \; \sqrt{\pi}+6 \; \lambda^2 \; \pi)}\nonumber\\ \nonumber\\
&&+\frac{2\sqrt{24 \; \alpha_0^4 \; \lambda^2 \; \pi-102 \; \alpha_0^3 \; 
\lambda^3 \pi^{3/2}+168 \; \alpha_0^2 \; \lambda^2 \; \pi^2 -126 \; \alpha_0 \; 
\lambda^5 \; \pi^{5/2}+36 \; \lambda^6 \; \pi^3}}{\pi(4 \; \alpha_0^2-9 \; 
\alpha_0 \; \lambda \; \sqrt{\pi}+6 \; \lambda^2 \; \pi)} \; , \nonumber \\
\end{eqnarray}
\noindent
is such that $C_0 \rightarrow -\alpha_0/{2\pi}$, $C_2 \rightarrow 0$ when 
$\lambda \rightarrow \infty$ and leads to the elimination of errors up to 
${\cal O}\left(\frac{k^2}{\lambda^2}\right)$.

This procedure can be systematically extended to higher orders, although with 
exponentially increasing algebraic complexity. It is important to observe that 
the scaling of the parameters, and hence the power-counting scheme, depends on 
the regularization procedure.

In Fig. 2 we plot the parameters obtained in the leading order and 
next-to-leading order calculation as a function of the cutoff. Note that when 
we add the term proportional to $C_2$ the parameter $C_0$ changes. The 
numerical evaluation here and in what follows is implemented with the 
underlying $D=1$ theory fixed by choosing $\alpha_0=2$, which corresponds to an 
exact binding energy $E_0=1$.

\begin{figure}
\centerline{\epsffile{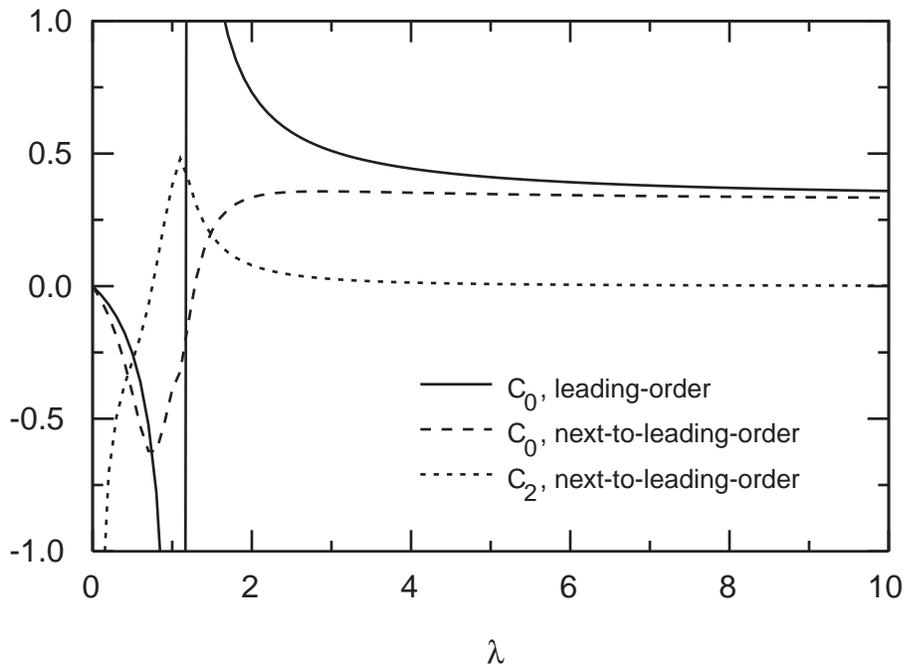}}
\vspace{0.5cm}
\caption{The leading-order and next-to-leading-order EFT expansion parameters 
for the one-dimensional delta-function potential.}
\end{figure}

We proceed by evaluating the errors in the binding energy, using the effective 
potential with one, two and four parameters adjusted to fit the inverse 
K-matrix. Instead of solving  systems of non-linear coupled equations, we fix 
the coefficients by applying the minimization procedure described in Section 2, 
which is simpler numerically and leads to the same results.

We solve the Schr\"{o}dinger equation numerically for different values of the 
cutoff $\lambda$. First, we introduce an integration cutoff $\Lambda_0 >> 
\lambda$, such that 
\begin{equation}
p^2 \Phi_{\lambda}(p)+\int_{-\Lambda_0}^{\Lambda_0} \; dq 
\;V_{\lambda}(p,p')\;\Phi_{\lambda}(q)=E_{\lambda} \Phi_{\lambda}(p) \; .
\end{equation}
\noindent
We discretize the Schr\"{o}dinger equation (choosing a gaussian quadrature)
\begin{equation}
\sum_{j=1}^{N}w_j \left[p_i p_j \; \delta_{ij}\; 
\frac{1}{w_j}+V_{\lambda}(p_i,p_j)\right]\; \Phi_{\lambda}(p_j)=E_{\lambda} \; 
\Phi_{\lambda}(p_i) \; ,
\end{equation}
\noindent
obtaining a matrix equation
\begin{equation}
({\bf M}-E_{\lambda}\; {\bf 1}){\bf \Phi_{\lambda}}=0 \; ,
\end{equation}
\noindent
where
\begin{equation}
M_{ij}=w_j \left[p_i p_j \; \delta_{ij}\; 
\frac{1}{w_j}+V_{\lambda}(p_i,p_j)\right]\; .
\end{equation}
\noindent
Next, the matrix ${\bf M}$ is symmetrized, reduced to tridiagonal form and 
the resulting matrix equation is solved by the QL method. The smallest 
(negative) eigenvalue corresponds to the bound-state energy. We then refine the 
result using Richardson extrapolation ~\cite{rich} in order to eliminate the 
dominant errors due to the integration cutoff, which are proportional to 
$E_0/{\Lambda_0}$. Removing these errors should not be confused with
removing $\lambda$-dependent errors. Finally, we compare the resulting
bound-state energy, 
$E_{\lambda}$, with the exact value.

In Fig. 3 we show in a log-log plot the absolute values for the relative
errors  in the bound-state energy as functions of $E_0/{\lambda^2}$. As
expected, we  obtain straight lines with slope given by the dominant power of 
$E_0/{\lambda^2}$ in the error, in agreement with the scaling analysis. 

\begin{figure}
\centerline{\epsffile{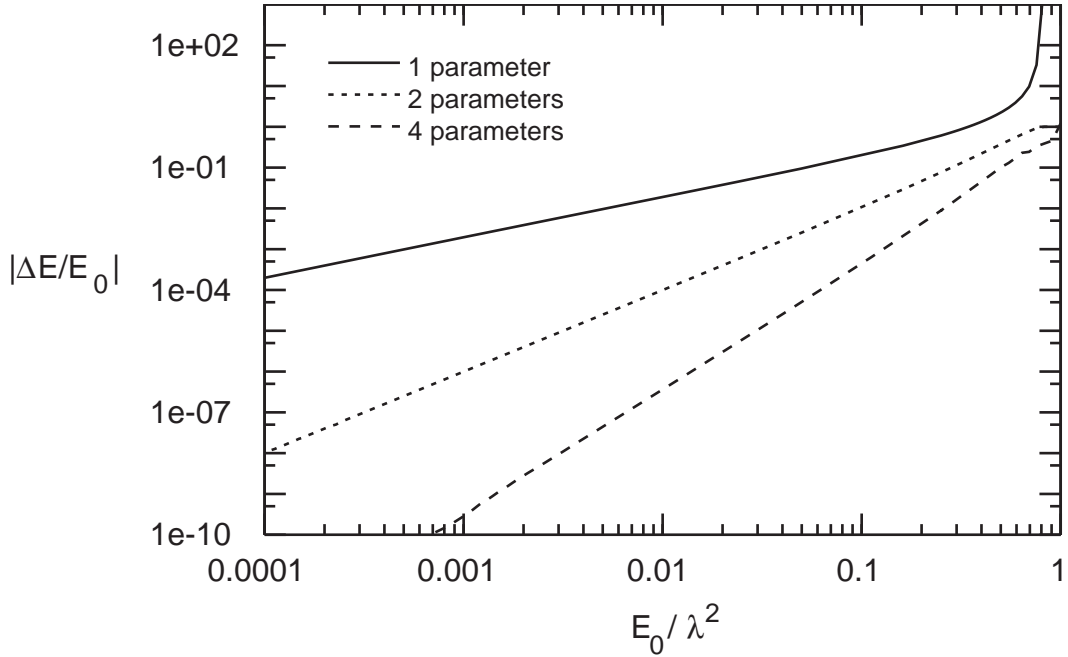}}
\vspace{0.5cm}
\caption{The EFT error in the binding energy for the one-dimensional 
delta-function using one, two and four parameters. The exact theory is fixed by 
choosing $\alpha_0=2$.}
\end{figure}

This result clearly demonstrates the systematic improvement in the accuracy of 
the effective potential for describing the exact theory in the low-energy 
regime. We also note that the errors corresponding to each approximation become 
comparable as $\lambda^2$ approaches the binding energy, indicating the 
breakdown of the effective theory. In this way, the error plot provides a 
graphical illustration of the radius of convergence, the point where all the 
error lines converge. 
%

%

%

\subsection{Similarity Renormalization Group Approach}

In the one-dimensional case the canonical hamiltonian in momentum space with a 
delta-function potential can be written as $H(p,p')=h(p,p')+V(p,p')$, where 
$h(p,p')=p^2 \delta^{(1)}(p-p')$ corresponds to the free hamiltonian and 
$V(p,p')=-{\alpha_0}/2\pi$ corresponds to the Fourier transform of the 
delta-function potential. This hamiltonian does not produce divergences so an 
initial cutoff is not required to define the problem.

The flow equation obtained with Wegner's transformation in terms of matrix 
elements in the basis of free states is given by 
\begin{equation}
\frac{dV_s(p,p')}{ds}=-(p^2-p'^2)^2 \; V_{s}(p,p')-\int_{-\infty}^{\infty}dk \; 
(2 k^2-p^2-p'^2)\; V_{s}(p,k)\; V_{s}(k,p') \; ,
\end{equation}
\noindent
with the initial condition $H_{s=0}(p,p')=H(p,p')$.
\noindent
The reduced interaction ${\bar V}_{s}(p,p')$ is defined such that
\begin{equation}
V_{s}(p,p')=e^{-s(p^2-p'^2)^2}\; {\bar V}_{s}(p,p') \; .
\end{equation}
\noindent
Now, assuming that $h$ is cutoff independent we obtain the flow equation for 
the reduced interaction, 
\begin{eqnarray}
\frac{d{\bar V}_{s}}{ds}=-e^{-2s\; p^2  p'^2}\int_{-\infty}^{\infty}&&dk \; (2 
k^2-p^2-p'^2)\; e^{-2s[k^4-k^2(p^2+p'^2)]}\nonumber\\
&&\times {\bar V}_{s}(p,k)\; {\bar V}_{s}(k,p') \; .
\label{fl}
\end{eqnarray}

In order to solve this equation, we employ a perturbative expansion
\begin{equation}
{\bar V}_{s}(p,p')={\bar V}^{(1)}_{s}(p,p')+{\bar V}^{(2)}_{s}(p,p')+ \cdots \; 
,
\label{pertans}
\end{equation}
\noindent
starting with
\begin{equation}
{\bar V}^{(1)}_{s}(p,p')=-\frac{{\alpha}_s}{2\pi} \; .
\end{equation}
Using coupling-coherence \cite{coupcoh} we assume a solution in the form of an 
expansion in powers of $\alpha_s/2\pi$,
\begin{equation}
{\bar 
V}_{s}(p,p')=-\frac{\alpha_s}{2\pi}+\sum_{n=2}^{\infty}\left(\frac{\alpha_{s}}{
2\pi}\right)^{n}\; F^{(n)}_{s}(p,p') \; ,
\label{coc}
\end{equation}
\noindent
with the additional condition that the operators $F^{(n)}_{s}(p,p')$ vanish 
when $p=p'=0$ or, equivalently, when $s=0$. Using  the solution  Eq.
(\ref{coc})  in Eq. (\ref{fl}) we obtain:
\begin{eqnarray}
\frac{d{\bar V}_{s}}{ds}&=&-\frac{1}{2\pi}\; 
\frac{d{\alpha}_s}{ds}+\sum_{n=2}^{\infty}\frac{1}{(2\pi)^n}\left[n \; 
\alpha_{s}^{n-1}\; \frac{d{\alpha}_s}{ds}\; F^{(n)}_{s}(p,p')+\alpha_{s}^{n}\; 
\frac{dF^{(n)}_{s}(p,p')}{ds}\right]\nonumber\\
&=&\int_{-\infty}^{\infty}dk \; (2 k^2-p^2-p'^2)\; e^{-2s[p^2 
p'^2+k^4-k^2(p^2+p'^2)]}\nonumber\\
&\times&\left[-\frac{\alpha_s}{2\pi}+\sum_{n=2}^{\infty}\left(\frac{\alpha_{s}}
{2\pi}\right)^{n}\; 
F^{(n)}_{s}(p,k)\right]\left[-\frac{\alpha_s}{2\pi}+\sum_{m=2}^{\infty}
\left(\frac{\alpha_{s}}{2\pi}\right)^{m}\; F^{(m)}_{s}(k,p')\right]\; .
\end{eqnarray}

This equation is solved iteratively order by order in $\alpha_s/2\pi$. If 
$\alpha_s/2\pi$ is small the operator ${\bar V}^{(1)}_{s}(p,p')$ can be 
identified as the dominant term in the expansion of ${\bar V}_{s}(p,p')$ in 
powers of $p$ and $p'$. In fact, since in the limit when $p$ and $p'$ approach 
zero all $F^{(n)}_{s}(p,p')$ vanish, the only term left is $-\alpha_s/(2 \pi)$, 
which in the $D=1$ case corresponds to a relevant operator (even though the 
coupling is dimensionless, it is multiplied by $2m=1$). The higher-order terms 
correspond to irrelevant operators. Thus, at each order $n$ the solution can be 
viewed as a double expansion obtained  by solving the equations for $\alpha_s$ 
and $F^{(n)}_{s}(p,p')$ separately.

At second-order we have
\begin{equation}
-\frac{1}{2 \pi}\; \frac{d{\alpha}_s}{ds}+\frac{1}{(2\pi)^2}\; 
\alpha_{s}^{2}\;\;  \frac{dF^{(2)}_{s}(p,p')}{ds}=- \alpha_s^2 \; 
I_{s}^{(2)}(p,p')\; ,
\label{v2}
\end{equation}
\noindent
where
\begin{equation}
I_{s}^{(2)}(p,p')=\frac{1}{(2\pi)^2}\; \int_{-\infty}^{\infty}dk \; (2 
k^2-p^2-p'^2)\; e^{-2s[p^2 p'^2+k^4-k^2(p^2+p'^2)]}\; .
\end{equation}
\noindent
The equation for the running coupling, $\alpha_s$, is obtained by taking the 
limit $(p,p') \rightarrow 0$: 
\begin{equation}
\frac{1}{2 \pi}\; \frac{d{\alpha}_s}{ds}= \alpha_s^2 \; I_{s}^{(2)}(0,0) \; ,
\label{alp2}
\end{equation}
\noindent
where
\begin{equation}
I_{s}^{(2)}(0,0)=\frac{1}{(2\pi)^2}\; \int_{-\infty}^{\infty}dk \; 2 k^2 e^{-2s 
k^4}=\frac{1}{(2\pi)^2}\; \frac{\Gamma(3/4)}{2^{3/4}}\; \frac{1}{s^{3/4}} \; .
\end{equation}
\noindent
Integrating Eq. (\ref{alp2}) we obtain:
\begin{eqnarray}
&&\alpha_{s,2}=\frac{\alpha_{s_0}}{1-\eta_2 \; \frac{\alpha_{s_0}}{2 
\pi}\;(s^{1/4}-s_{0}^{1/4})} \; ,
\label{ralp2}
\\ \nonumber\\
&&\eta_2= \frac{4\Gamma(3/4)}{2^{3/4}} \; .
\end{eqnarray}
\noindent
The extra subscript on $\alpha$ displays the order of the right-hand side of
Eq. (104). Thus, knowing the value of  $\alpha_{s_0}$ for a given $s_0$,  Eq. 
(\ref{ralp2}) can be used to obtain  the running coupling $\alpha_s$ for any 
value $s$. The choice  $s_0=0$ corresponds to $\alpha_{s_0}=\alpha_0$ (``bare 
coupling''), such that
\begin{equation}
\alpha_{s,2}=\frac{\alpha_{0}}{1-\eta_2\; \frac{\alpha_{0}}{2\pi}\;s^{1/4}}\; .
\end{equation}
\noindent
In terms of the cutoff $\lambda$ we obtain:
\begin{equation}
\alpha_{\lambda,2}=\frac{\alpha_{0}}{1-\eta_2 \; \frac{\alpha_{0}}{2 
\pi}\;\frac{1}{\lambda}} \; .
\end{equation} 

The equation for $F^{(2)}_{s}(p,p')$ is given by: 
\begin{equation}
\frac{1}{(2\pi)^2}\frac{dF^{(2)}_{s}(p,p')}{ds}=I_{s}^{(2)}(0,0)- 
I_{s}^{(2)}(p,p') \; .
\end{equation}
\noindent
Integrating we obtain:
\begin{eqnarray}
F^{(2)}_{s}(p,p')&=&(2\pi)^2\int_{0}^{s}ds' 
\left[I_{s'}^{(2)}(0,0)-I_{s'}^{(2)}(p,p')\right]\nonumber\\
&=&\int_{-\infty}^{\infty}dk \; \left[\frac{1}{k^2}\left(1-e^{-2 s 
k^4}\right)\right. \nonumber\\ 
&+&\left. \left[e^{-2 s(k^2-p^2)(k^2-p'^2)}-1\right]\; 
\left[\frac{1}{2(k^2-p^2)}+\frac{1}{2(k^2-p'^2)}\right]\right] \; .
\end{eqnarray}
\noindent
The integral over $k$ must be performed numerically. 

At third-order we have:
\begin{eqnarray}
-\frac{1}{2 \pi}\; \frac{d{\alpha}_s}{ds}+\frac{1}{(2\pi)^2}\; \alpha_{s}^{2}\; 
\frac{dF^{(2)}_{s}(p,p')}{ds}
&+&\frac{2}{(2\pi)^2} \; \alpha_{s} \; \frac{d\alpha_{s}}{ds}\; 
F^{(2)}_{s}(p,p')+ \frac{1}{(2\pi)^3}\; \alpha_{s}^{3} \; 
\frac{dF^{(3)}_{s}(p,p')}{ds}\nonumber\\
&=&- \alpha_s^2 \; I_{s}^{(2)}(p,p')+ \alpha_s^3 \; I_{s}^{(3)}(p,p') \; ,
\label{v3}
\end{eqnarray}
\noindent
where
\begin{eqnarray}
I_{s}^{(3)}(p,p')&=&\frac{1}{(2\pi)^3}\int_{-\infty}^{\infty}dk \; (2 
k^2-p^2-p'^2)\; e^{-2s[p^2 p'^2+k^4-k^2(p^2+p'^2)]} \nonumber\\
&\times&  \left[F^{(2)}_{s}(p,k)+F^{(2)}_{s}(k,p')\right] \; .
\end{eqnarray}
\noindent
Taking the limit $p,p' \rightarrow 0$ we obtain:
\begin{equation}
\frac{1}{2 \pi}\; \frac{d{\alpha}_s}{ds}= \alpha_s^2 \; I_{s}^{(2)}(0,0)- 
\alpha_s^3 \; I_{s}^{(3)}(0,0) \; .
\label{alp3}
\end{equation}
\noindent
where
\begin{eqnarray}
I_{s}^{(3)}(0,0)=\frac{1}{(2\pi)^3}\;\int_{-\infty}^{\infty}dk \; 2 k^2 e^{-2s 
k^4}\;  \left[F^{(2)}_{s}(0,k)+F^{(2)}_{s}(k,0)\right] \; .
\end{eqnarray}
\noindent
For dimensional reasons Eq. (\ref{alp3}) takes the form
\begin{equation}
\frac{d{\alpha}_s}{ds}= \frac{B_2}{s^{3/4}} \; \alpha_s^2 - 
\frac{B_3}{s^{1/2}}\;  \alpha_s^3 \; ,
\label{alp33}
\end{equation}
\noindent
where $B_2$ and $B_3$ can be obtained by evaluating $I_{s}^{(2)}(0,0)$ and 
$I_{s}^{(3)}(0,0)$ for $s=1$. 
\noindent
In terms of the cutoff $\lambda$ we obtain:
\begin{equation}
\frac{d{\alpha}_{\lambda}}{d\lambda}= \frac{4 }{\lambda^{2}} \; B_2 \; 
\alpha_{\lambda}^{2} - \frac{4 }{\lambda^3}\; B_3 \; \alpha_{\lambda}^{3} \; .
\end{equation}
\noindent
Eq. (\ref{alp3}) leads to a transcendental equation which is solved
numerically  in order to obtain the running coupling $\alpha_{s,3}$.

A simple scaling analysis shows that the similarity function $f_{_{\lambda 
ij}}=e^{-\Delta_{ij}^2/\lambda^2}$ introduces errors that are expected to scale 
as inverse powers of $\lambda$. The leading-order errors are proportional to 
$\sqrt{E_0}/{\lambda}$ and are expected to be eliminated by using the running 
coupling renormalized to second-order. The next-to-leading-order errors are 
proportional to $E_0/{\lambda^2}$ and are expected to be eliminated by using 
the running coupling renormalized to third-order and including the second-order 
effective interaction, since both introduce corrections proportional to 
$E_0/{\lambda^2}$, which therefore enter at the same level. Thus, to show the 
systematic improvement in the perturbative approximation in our study of the 
one-dimensional delta-function we compute  the running coupling to third-order 
and the operators to second-order. As in the EFT calculation, we choose the 
exact binding-energy $E_0=1$ ($\alpha_0=2$) and calculate the bound-state 
energy in a given approximation by solving the Schr\"{o}dinger equation 
numerically. In Fig. 4 we show the running coupling renormalized to second and 
third-order as a function of the cutoff $\lambda$. In Fig. 5 we show in a 
log-log plot the absolute value of the relative errors in the binding-energy as 
a function of the ratio $E^{(0)}/{\lambda^2}$ using the following 
approximations for the interaction:

\vspace{0.5cm}

\noindent
(a) relevant operator with ``bare" coupling $\alpha_0$,
\begin{equation}
V_{\lambda}(p,p')=-\frac{\alpha_0}{2\pi}\; e^{-\frac{(p^2-p'^2)^2}{\lambda^4}} 
\; ;
\end{equation}

\noindent
(b) relevant operator with running coupling renormalized to second-order ( 
$\alpha_{\lambda,2})$,
\begin{equation}
V_{\lambda}(p,p')=-\frac{\alpha_{\lambda,2}}{2\pi}\; 
e^{-\frac{(p^2-p'^2)^2}{\lambda^4}} \; ;
\end{equation}

\noindent
(c) relevant operator plus second-order irrelevant operator with running 
coupling renormalized to third-order ($\alpha_{\lambda,3}, F^{(2)}_{\lambda}$),

\begin{equation}
V_{\lambda}(p,p')=\left[-\frac{\alpha_{\lambda,2}}{2\pi}+\left(\frac{\alpha_{\l
ambda,2}}{2\pi}\right)^2 \; F^{(2)}_{\lambda}(p,p')\right] \; 
e^{-\frac{(p^2-p'^2)^2}{\lambda^4}} \; ;
\end{equation}

\noindent
(d) relevant operator with running coupling renormalized to third-order  
($\alpha_{\lambda,3}$),
\begin{equation}
V_{\lambda}(p,p')=-\frac{\alpha_{\lambda,3}}{2\pi}\; 
e^{-\frac{(p^2-p'^2)^2}{\lambda^4}} \; ;
\end{equation}

\noindent
(e) relevant operator plus second-order irrelevant operator with running 
coupling renormalized to third-order, ($\alpha_{\lambda,3}, 
F^{(2)}_{\lambda})$,
\begin{equation}
V_{\lambda}(p,p')=\left[-\frac{\alpha_{\lambda,3}}{2\pi}+\left(\frac{\alpha_{\l
ambda,3}}{2\pi}\right)^2 \; F^{(2)}_{\lambda}(p,p')\right]  \; 
e^{-\frac{(p^2-p'^2)^2}{\lambda^4}}  \; .
\end{equation}
\vspace{0.5cm}

\begin{figure}
\centerline{\epsffile{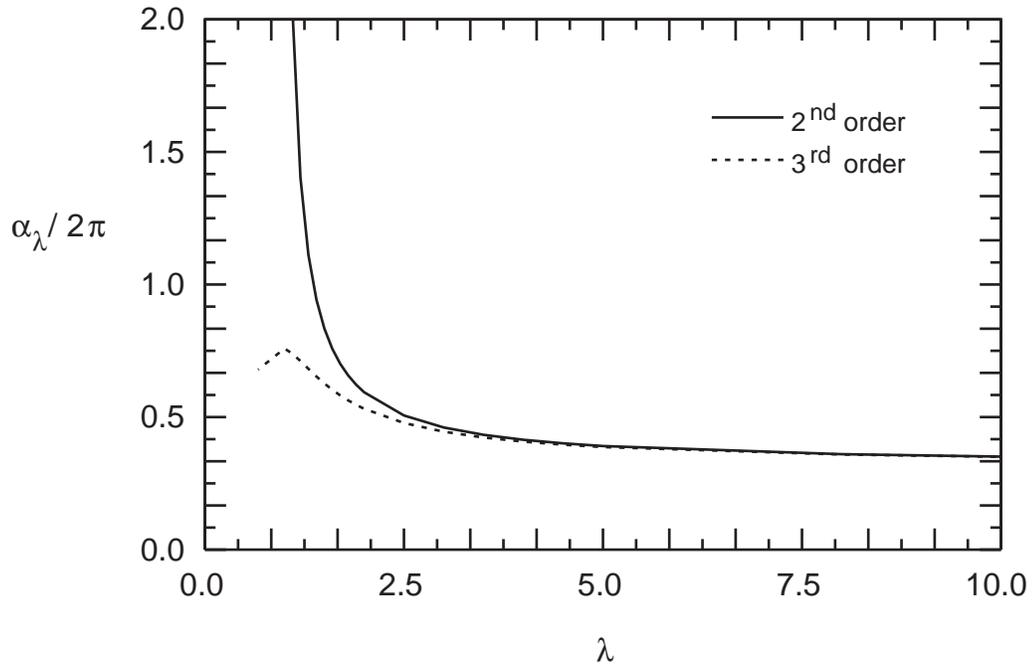}}
\vspace{0.5cm}
\caption{The SRG running coupling for the one-dimensional delta-function.}
\end{figure}

\begin{figure}
\centerline{\epsffile{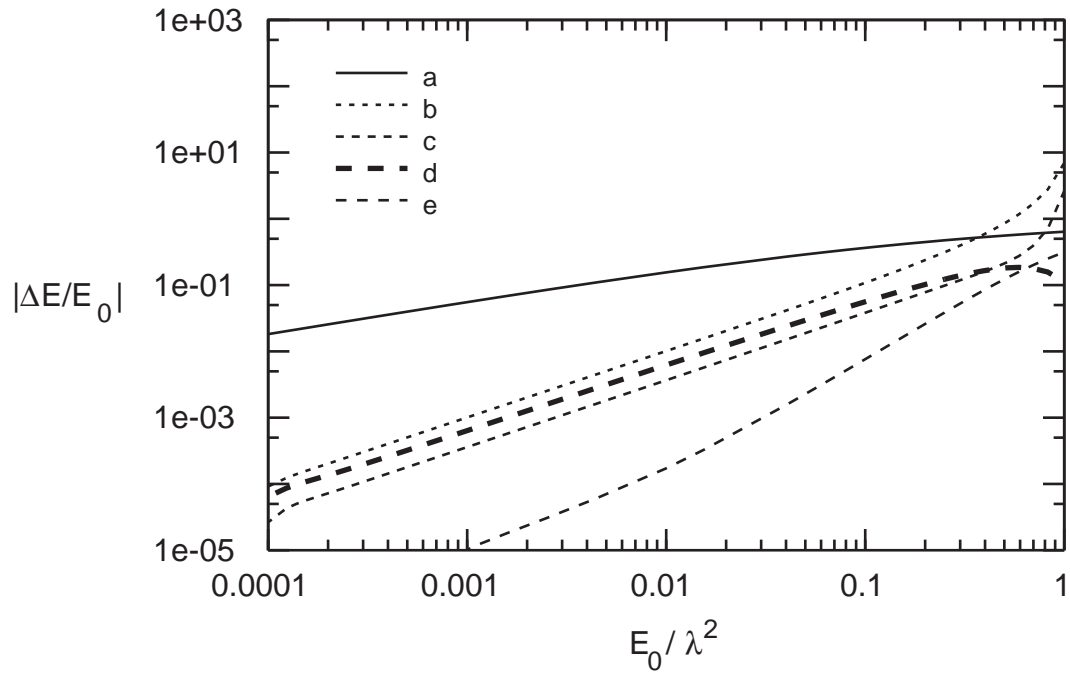}}
\vspace{0.5cm}
\caption{The SRG error in the binding energy for the one-dimensional 
delta-function using various approximations for the SRG hamiltonian. The exact 
theory is fixed by choosing $\alpha_0=2$.}
\end{figure}

With hamiltonian (a) (unrenormalized) the dominant error is proportional 
$\sqrt{E_0}/{\lambda}$, which is the leading-order error introduced by the 
regulator. With the renormalized hamiltonians (b), (c), (d) and (e), for 
intermediate values of $\lambda$, we obtain straight lines with slope given by 
the dominant power of $E_0/{\lambda^2}$ in the error. With hamiltonian (b) the 
slope is one, corresponding to an error proportional to  $E_0/{\lambda^2}$. 
With hamiltonians (c) and (d) there is only a small shift in the errors; there 
is no significant change in the slope. With hamiltonian (e) the slope is two, 
corresponding to an error proportional to $(E_0/{\lambda^2})^2$. Analogous to 
the EFT calculation, for small values of $\lambda$ the lines converge, 
indicating the breakdown of the perturbative expansion. It is important to note 
that for very large values of the cutoff $\lambda$ the scaling behavior might 
change, since higher-order contributions not included in a given approximation 
for the hamiltonian can become important. In our calculations we did not 
analyze this regime because numerical accuracy limits the maximum range of 
cutoffs on the integrals.

\newpage
%

%

\section{Two-Dimensional Delta-function Potential}

We now consider the case of two nonrelativistic particles in two dimensions 
interacting via an attractive  Dirac delta-function potential. The  
Schr\"{o}dinger equation for relative motion in position space (with
$\hbar=1$), can be written as:
\begin{equation}
-{\bf \nabla}_{\bf r}^2 \Psi({\bf r})-\alpha_0 \;\delta^{(2)}({\bf r}) \; 
\Psi({\bf r})=E \; \Psi({\bf r})\; .
\end{equation}
\noindent
Both the delta-function potential in two dimensions and the kinetic energy 
operator  scale as $1/r^2$, therefore, the coupling  $\alpha_0$ is 
dimensionless. As a consequence, the hamiltonian is scale invariant ({\it
i.e.},  there is no intrinsic energy scale) and we can anticipate the presence
of   logarithmic ultraviolet divergences, analogous to those appearing in QED
and QCD. The problem requires  renormalization. In this subsection we present
the standard method that  produces an exact solution analytically, using simple
regularization and  renormalization schemes \cite{jackiw}.

\subsection{Exact Solution}

We start with the Schr\"{o}dinger equation in momentum space,
\begin{equation}
p^2 \;  \Phi({\bf p})-\frac{\alpha_0}{(2\pi)^2}\; \int \; d^{2}q \;\;\Phi({\bf 
q})=E \; \Phi({\bf p}) \; ,
\label{semom2}
\end{equation}
\noindent
where $\Phi({\bf p})$ is the Fourier transform of the position space 
wave-function,
\begin{equation}
\Phi({\bf p})=\frac{1}{{2\pi}}\int \; d^{2}r  \;\; \Psi({\bf r})\;e^{-i {\bf 
p}.{\bf r}} \; .
\end{equation}

As a consequence of scale invariance, if there is any negative energy solution
to  Eq. (\ref{semom2}) then it will admit solutions for any $E<0$. This 
corresponds  to a continuum of bound states with energies extending down to
$-\infty$, so  the system is not bounded from below. By rearranging the terms
in the  Schr\"{o}dinger equation we obtain   
\begin{equation}
\Phi({\bf p})=\frac{\alpha_0}{2\pi}\; \frac{\Psi(0)}{(p^2+E_0)} \; ,
\label{2ds}
\end{equation}
\noindent
where $\Psi(0)$ is the position space wave-function at the origin and $E_0 >0$ 
is the binding energy.

To obtain the eigenvalue condition for the binding energy, we can integrate 
both sides of Eq.~(\ref{2ds}):
\begin{equation}
1=\frac{\alpha_0}{2\pi}\;  \int_{0}^{\infty} \; dp \; p \; \frac{1}{(p^2+E_0)} 
\; .
\label{bsi2}
\end{equation}
\noindent
The integral on the r.h.s. diverges logarithmically, so the problem is 
ill-defined. 

The conventional way to deal with this problem is renormalization. First, we 
regulate the integral with a momentum cutoff, obtaining
\begin{equation}
1=\frac{\alpha_0}{2\pi}\;  \int_{0}^{\Lambda} \; dp \; p \; 
\frac{1}{(p^2+E_0)}=\frac{\alpha_0}{4\pi}\; {\rm 
ln}\left(1+\frac{\Lambda^2}{E_0}\right) \; ,
\end{equation}
\noindent
so that 
\begin{equation}
E_0=  \frac{\Lambda^2}{e^{-\frac{4\pi}{\alpha_0}}-1} \; .
\end{equation}

Clearly, if the coupling $\alpha_0$ is fixed then $E_0 \rightarrow \infty$ as 
$\Lambda \rightarrow \infty$. In order to eliminate the divergence and produce 
a finite, well-defined bound state we can renormalize the theory by demanding 
that the coupling runs with the cutoff $\Lambda$ in such a way that the binding 
energy remains fixed as the cutoff is removed:
\begin{equation}
\alpha_0 \rightarrow \alpha_{\Lambda}=\frac{4\pi}{{\rm 
ln}\left(1+\frac{\Lambda^2}{E_0}\right)} \; .
\end{equation}

The dimensionless renormalized running coupling $\alpha_{\Lambda}$  that 
characterizes the strength of the interaction is therefore replaced by a new 
(dimensionful) parameter $E_0 >0$, the binding energy of the system. This is a 
simple example of dimensional transmutation: even though the original ``bare'' 
hamiltonian is scale invariant, the renormalization procedure leads to a scale 
that characterizes the physical observables. Note that $E_0$ can be chosen 
arbitrarily, fixing the energy scale of the underlying (renormalized) theory. It 
is also interesting to note that the renormalized running coupling 
$\alpha_{\Lambda}$ vanishes as $\Lambda \rightarrow \infty$ and so the theory 
is asymptotically free.

This renormalized hamiltonian can be used to compute other observables. The 
usual prescription for the calculations is to obtain the solutions with the 
cutoff in place and then take the limit as the momentum cutoff is removed to 
$\infty$. If an exact calculation can be implemented, the final results should 
be independent of the regularization and renormalization schemes. As an 
example, we calculate the scattering wave function,
\begin{equation}
\Phi_{k}({\bf p})=\delta^{(2)}({\bf p}-{\bf k})+\frac{\alpha_{\Lambda}}{2 
\pi}\; \frac{\Psi(0)}{(p^2-k^2-i \; \epsilon)} \; ,
\end{equation}
\noindent
where $k=\sqrt{E}$.
\noindent
Integrating both sides over ${\bf p}$ with a cutoff $\Lambda$ in place, we 
obtain
\begin{equation}
\Psi(0)=\frac{1}{2\pi} \; \left[1-\frac{\alpha_{\Lambda}}{4\pi}\; {\rm 
ln}\left(1+\frac{\Lambda^2}{-k^2-i\epsilon}\right)\right]^{-1} \; ;
\end{equation}
\noindent
thus,
\begin{equation}
\alpha_{\Lambda} \; \Psi(0)=\frac{1}{2\pi} \; \left[\frac{1}{4\pi}\;{\rm 
ln}\left(1+\frac{\Lambda^2}{E_0}\right)-\frac{1}{4\pi}\; {\rm 
ln}\left(1+\frac{\Lambda^2}{-k^2-i\epsilon}\right)\right]^{-1} \; .
\end{equation}
\noindent
In the limit $\Lambda \rightarrow \infty$ we obtain:
\begin{equation}
\alpha_{\Lambda} \; \Psi(0)=\frac{2}{{\rm ln}\left(\frac{k^2}{E_0}\right)-i \; 
\pi} \; .
\end{equation}
\noindent
The resulting scattering wave function is then given by
\begin{equation}
\Phi_{k}({\bf p})=\delta^{(2)}({\bf p}-{\bf k})+\frac{1}{2 \pi}\; 
\frac{2}{(p^2-k^2-i\epsilon)}\left[{\rm ln}\left(\frac{k^2}{E_0}\right)-i \; 
\pi\right]^{-1} \; .
\end{equation}

It is important to note that only S-wave scattering occurs, corresponding to 
zero angular momentum states. For the higher waves  the centrifugal barrier 
completely screens the delta-function potential and the non-zero angular 
momentum scattering states are free states. 

The same prescription  can be used to  evaluate the T-matrix or the K-matrix.
For the T-matrix,  the Lippmann-Schwinger equation with the renormalized 
potential is given by:
\begin{equation}
T({\bf p},{\bf p'};k)=V({\bf p},{\bf p'})+\int \; d^{2}q \; \; \frac{V({\bf 
p},{\bf q})}{k^2-q^2+i \epsilon} \; T({\bf q},{\bf p'};k) \; .
\end{equation}
\noindent
Since only S-wave scattering takes place we can integrate out the angular 
variable, obtaining 
\begin{equation}
T^{(\rm l=0)}( p,p';k)=V^{(\rm l=0)}(p,p')+\int_{0}^{\Lambda} \; dq \; q \;  
\frac{V^{(\rm l=0)}(p,q)}{k^2-q^2+i \epsilon} \; T^{(\rm l=0)}(q,p';k) \; ,
\end{equation}
\noindent
where
\begin{equation}
V^{(\rm l=0)}(p,p')=-\frac{\alpha_{\Lambda}}{2\pi} \; .
\end{equation}

The Lippmann-Schwinger equation for the ``on-shell'' T-matrix is given by:
\begin{eqnarray}
T^{(\rm l=0)}(k)=-\frac{\alpha_{\Lambda}}{2\pi}-\frac{\alpha_{\Lambda}}{2\pi}\; 
T^{(\rm l=0)}(k) \; \int_{0}^{\Lambda} \; dq \; q \;  \frac{1}{k^2-q^2+i 
\epsilon}.
\end{eqnarray}
\noindent
Solving this equation and taking the limit $\Lambda \rightarrow \infty$, we 
obtain the exact ``on-shell'' T-matrix:
\begin{equation}
T_{0}(k)= -\frac{2}{{\rm ln}\left(\frac{k^2}{E_0}\right)-i \; \pi}\; .
\end{equation}
\noindent
Here and in what follows we drop the superscript and use the subscript $0$ to 
denote the exact quantities.

In the same way, the S-wave Lippmann-Schwinger equation for the K-matrix is 
given by
\begin{equation}
K(p,p';k)=V(p,p')+{\cal P}\int_{0}^{\Lambda} \; dq \; q \; 
\frac{V(p,q)}{k^2-q^2} \; K(q,p';k) \; ,
\end{equation}
\noindent
and the exact ``on-shell'' K-matrix is given by
\begin{equation}
K_0(k)= -\frac{2}{{\rm ln}\left(\frac{k^2}{E_0}\right)}\; .
\end{equation}

The ``on-shell'' K-matrix and T-matrix  are related by
\begin{equation}
K_0(k)=\frac{T_0(k)}{1-\frac{i\pi}{2}T_0(k)} \; .
\end{equation}
\noindent
Using either 
\begin{equation}
k \; {\rm cot}\;\delta_{0}(k)-i k=-\frac{2 \; k}{\pi}\frac{1}{T_0(k)} \; ,
\end{equation}
\noindent
or
\begin{equation}
k \;{\rm cot}\; \delta_0(k)=-\frac{2 \; k}{\pi}\frac{1}{K_0(k)} \; ,
\end{equation}
\noindent
we can obtain the exact phase-shifts:
\begin{equation}
{\rm cot} \; \delta_{0}=\frac{1}{\pi}\; {\rm ln}\left(\frac{k^2}{E_0}\right)\; 
.
\end{equation}
%

%

%

\subsection{Effective Field Theory Approach}

The EFT power counting scheme in the $D=2$ case is also simple. Again, we 
analyze the errors in the inverse K-matrix. For the leading-order prediction, 
we use the effective potential with one parameter. Evaluating the difference
$\Delta \; [1/K^{(0)}]=1/K_{\lambda}^{(0)}-1/K_0$ and expanding it in powers of 
$k/{\lambda}$ we obtain:
\begin{equation}
\Delta \; \left[\frac{1}{K^{(0)}}\right]=
\left[\frac{1}{C_0}-\frac{\gamma}{2}+\frac{1}{2}\; {\rm 
ln}\left(\frac{\lambda^2}{E_0}\right)\right]
+\left[\frac{1}{C_0}-\frac{1}{2}\right]\; \frac{k^2}{\lambda^2}+{\cal 
O}\left(\frac{k^4}{\lambda^4}\right)\; .
\end{equation}
By choosing 
\begin{equation}
C_0(\lambda)=\frac{-2}{{\rm ln}\left(\frac{\lambda^2}{E_0}\right)-\gamma}
 \; ,
\end{equation}
\noindent
the leading logarithmic error is eliminated and the remaining errors are 
dominated by the term 
\begin{equation}
{\cal O}\left(\frac{k^2}{\lambda^2}\right)
=\frac{1}{2}\left[\gamma-{\rm 
ln}\left(\frac{\lambda^2}{E_0}\right)-1\right]\frac{k^2}{\lambda^2}\; .
\end{equation}

For the next-to-leading order calculation we use the effective potential with 
two parameters, obtaining the expansion
\begin{equation}
\Delta \; \left[\frac{1}{K^{(2)}}\right]=A_0+A_2 \; \frac{k^2}{\lambda^2}+{\cal 
O}\left(\frac{k^4}{\lambda^4}\right)\; ,
\end{equation}
\noindent
where
\begin{eqnarray}
A_0&=&\frac{4+2\; C_2+\frac{C_2^2}{4}}{2\left(2 \; C_0-\frac{C_2^2}{4}\right)}
-\frac{1}{2}\; \left[\gamma-{\rm ln}\left(\frac{\lambda^2}{E_0}\right)\right] 
\; ,
\\ \nonumber\\
A_2&=&-\frac{4\;C_0^2-2\; C_0\;(4+2\;C_2+\frac{3\;C_2^2}{4})+C_2\;(8+12\; C_2 
+3\;C_2^2+\frac{3\;C_2^3}{8})}{2\; (2\;C_0-\frac{C_2^2}{4})^2}\; .
\end{eqnarray}
\noindent
The parameters $C_0$ and $C_2$ are determined by solving the system of two 
coupled non-linear equations given by
\begin{equation}
A_0=0 \; , \; \; A_2=0 \; .
\end{equation}
\noindent
The acceptable solution is the one that leads to asymptotic freedom,
{\it i.e.}, $(C_0, C_2) \rightarrow 0$ as $\lambda \rightarrow \infty$ (for
the other  solution, $C_0 \rightarrow 8$ and $C_2 \rightarrow 0$ in this
limit) . The  corresponding expressions for $C_0$ and $C_2$ are algebraically
very  complicated and so we do not present them here. In Fig. 6 we plot the 
parameters obtained in the leading order and next-to-leading order calculation 
as a function of the cutoff. For the numerical evaluation here and in what 
follows we fix the $D=2$ theory by choosing the exact binding energy to be 
$E_0=1$.

\begin{figure}
\centerline{\epsffile{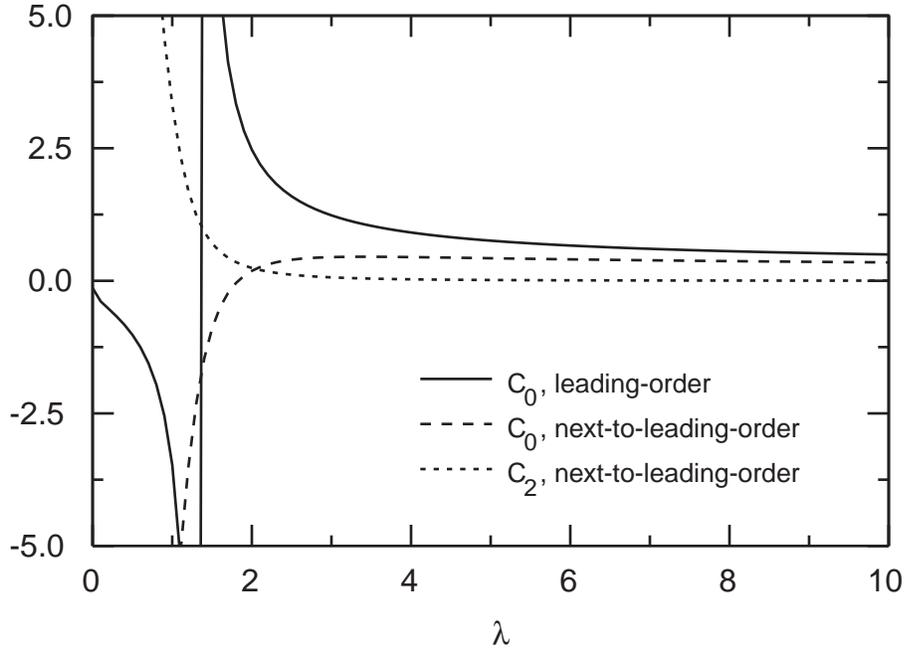}}
\vspace{0.5cm}
\caption{The leading-order and next-to-leading-order EFT expansion parameters 
for the two-dimensional delta-function potential.}
\end{figure}

To evaluate the errors in the binding energy we use the effective potential 
with one, two and four parameters adjusted to fit the inverse K-matrix (using 
the same minimization procedure as in the $D=1$ case ).

We solve the Schr\"{o}dinger equation numerically for different values of the 
cutoff $\lambda$. Introducing an integration cutoff $\Lambda_0 >> \lambda$, 
\begin{equation}
p^2 \Phi_{\lambda}(p)+\int_{0}^{\Lambda_0} \; dq \; q 
\;V_{\lambda}(p,p')\;\Phi_{\lambda}(q)=E_{\lambda} \Phi_{\lambda}(p) \; .
\end{equation}
\noindent
Discretizing the Schr\"{o}dinger equation (choosing a gaussian quadrature),
\begin{equation}
\sum_{j=1}^{N}w_j \left[p_i p_j \; \delta_{ij}\; \frac{1}{w_j}+p_j \; 
V_{\lambda}(p_i,p_j)\right]\; \Phi_{\lambda}(p_j)=E_{\lambda} \; 
\Phi_{\lambda}(p_i) \; ,
\end{equation}
\noindent
we obtain the matrix equation
\begin{equation}
({\bf M}-E_{\lambda}\; {\bf 1}){\bf \Phi_{\lambda}}=0 \; ,
\end{equation}
\noindent
where
\begin{equation}
M_{ij}=w_j \left[p_i p_j \; \delta_{ij}\; \frac{1}{w_j}+p_j \; 
V_{\lambda}(p_i,p_j)\right]\; .
\end{equation}
\noindent
The bound-state energy, $E_{\lambda}$, is obtained by following the same 
procedure as for the $D=1$ case and  is compared with the exact value.

In Fig. 7 we show in a log-log plot the absolute values for the relative
errors  in the bound-state energy as a function of $E_0/{\lambda^2}$. As
expected, the  result is analogous to the $D=1$ case: straight lines with
slope given by the  dominant power of $E_0/{\lambda^2}$ in the error. Note
that the dominant  logarithmic errors are completely absorbed by adjusting the
parameter $C_0$ and   when more parameters are adjusted we obtain power-law
improvement.

\begin{figure}
\centerline{\epsffile{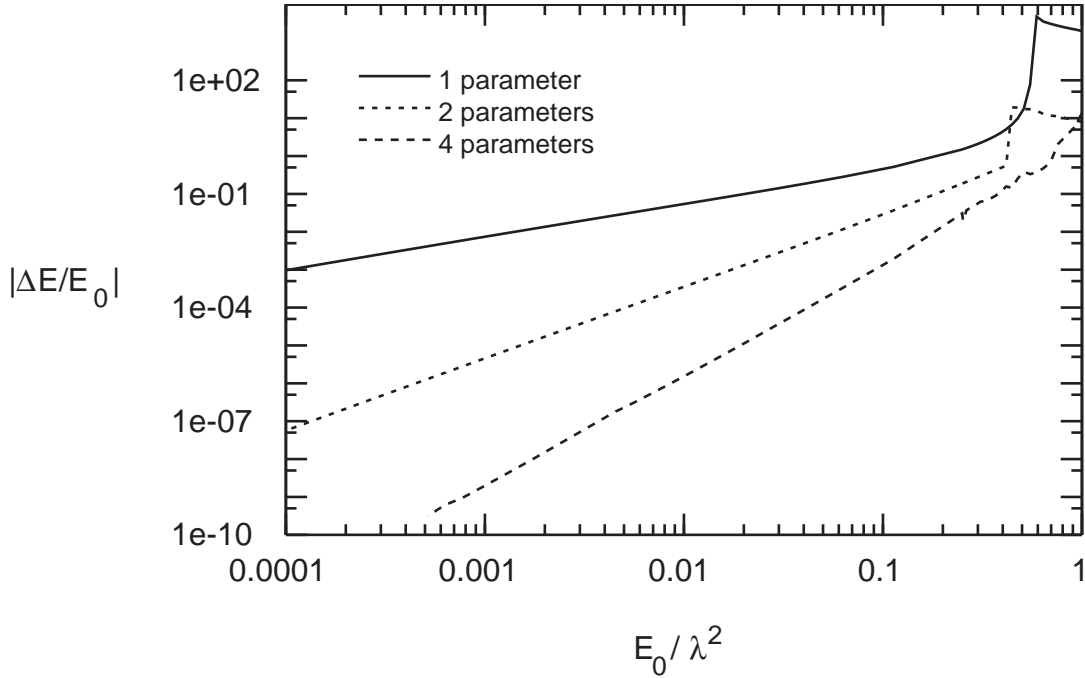}}
\vspace{0.5cm}
\caption{The EFT error in the binding energy for the two-dimensional 
delta-function using one, two and four parameters. The exact theory is fixed by 
choosing $E_0=1$.}
\end{figure}
%

%

%

\subsection{Similarity Renormalization Group Approach}

In the two-dimensional case the canonical hamiltonian in momentum space with a 
delta-function potential can be written as 
\begin{equation}
H({\bf p},{\bf p'})=h({\bf p},{\bf p'})+V({\bf p},{\bf p'}) \; , 
\end{equation}
\noindent
where $h({\bf p},{\bf p'})=p^2 \delta^{(2)}({\bf p}-{\bf p'})$ corresponds to 
the free hamiltonian and $V({\bf p},{\bf p'})=-{\alpha_0}/(2\pi)^2$ corresponds 
to the Fourier transform of the delta-function potential.

Integrating out the angular variable, the flow equation obtained with Wegner's 
transformation in terms of matrix elements in the basis of free states is given 
by 
\begin{equation}
\frac{dV_s(p,p')}{ds}=-(p^2-p'^2)^2 \; V_{s}(p,p')-\int_{0}^{\infty}dk \; k \; 
(2 k^2-p^2-p'^2)\; V_{s}(p,k)\; V_{s}(k,p') \; .
\end{equation}
\noindent
In principle, we can set the boundary condition at $s=0$ (no cutoff), i.e, 
\begin{equation}
H_{s=0}(p,p')=H(p,p')=p^2 \delta^{(1)}(p-p')-\frac{\alpha_0}{2\pi} \; .
\end{equation}
\noindent
However, the hamiltonian with no cutoff produces logarithmic divergences, 
requiring renormalization. As we will see, the boundary condition must be 
imposed at some other point, leading to dimensional
transmutation~\cite{coleman}.
\noindent
The reduced interaction ${\bar V}_{s}(p,p')$ is defined such that
\begin{equation}
V_{s}(p,p')=e^{-s(p^2-p'^2)^2}\; {\bar V}_{s}(p,p') \; .
\end{equation}
\noindent
Assuming that $h$ is cutoff independent we obtain the flow equation for the 
reduced interaction, 
\begin{eqnarray}
\frac{d{\bar V}_{s}}{ds}=-e^{-2s\; p^2  p'^2}\; \int_{0}^{\infty}&&dk \; k \; 
(2 k^2-p^2-p'^2)\; e^{-2s[k^4-k^2(p^2+p'^2)]}\nonumber\\
&&\times {\bar V}_{s}(p,k)\; {\bar V}_{s}(k,p') \; .
\label{f2}
\end{eqnarray}

As in the $D=1$ case, this equation is solved using  a perturbative expansion  
given by Eq. (~\ref{pertans}), starting with
\begin{equation}
{\bar V}^{(1)}_{s}(p,p')=-\frac{{\alpha}_s}{2\pi} \; .
\end{equation}
\noindent
We assume a coupling-coherent solution in the form of an expansion in powers of 
$\alpha_s/2\pi$, satisfying the constraint that the operators 
$F^{(n)}_{s}(p,p')$ vanish when $p=p'=0$,
\begin{equation}
{\bar 
V}_{s}(p,p')=-\frac{\alpha_s}{2\pi}+\sum_{n=2}^{\infty}\left(\frac{\alpha_{s}}{
2\pi}\right)^{n}\; F^{(n)}_{s}(p,p') \; .
\label{coc2}
\end{equation}
\noindent
Note that the expansion parameter is $\alpha_s/2\pi$.

Using  the solution  Eq. (\ref{coc2}) in Eq. (\ref{f2}) we obtain
\begin{eqnarray}
\frac{d{\bar V}_{s}}{ds}&=&-\frac{1}{(2\pi)^2}\; 
\frac{d{\alpha}_s}{ds}+\sum_{n=2}^{\infty}\frac{1}{(2\pi)^n}\left[n \; 
\alpha_{s}^{n-1}\; \frac{d{\alpha}_s}{ds}\; F^{(n)}_{s}(p,p')+\alpha_{s}^{n}\; 
\frac{dF^{(n)}_{s}(p,p')}{ds}\right]\nonumber\\
&=&\int_{0}^{\infty}dk \; k \; (2 k^2-p^2-p'^2)\; e^{-2s[p^2 
p'^2+k^4-k^2(p^2+p'^2)]}\nonumber\\
&\times&\left[-\frac{\alpha_s}{2\pi}+\sum_{n=2}^{\infty}\left(\frac{\alpha_{s}}
{2\pi}\right)^{n}\; 
F^{(n)}_{s}(p,k)\right]\left[-\frac{\alpha_s}{2\pi}+\sum_{m=2}^{\infty}
\left(\frac{\alpha_{s}}{2\pi}\right)^{m}\;
F^{(m)}_{s}(k,p')\right] \; .
\end{eqnarray}
\noindent
This equation is solved iteratively order-by-order in $\alpha_s/2\pi$. Again, 
if $\alpha_s/2\pi$ is small the operator ${\bar V}^{(1)}_{s}(p,p')$ can be 
identified as the dominant term in the expansion of ${\bar V}_{s}(p,p')$ in 
powers of $p$ and $p'$. In the $D=2$ case this operator corresponds to a 
marginal operator (since the coupling is dimensionless and there is no implicit 
mass scale). The higher-order terms correspond to irrelevant operators. 

At second-order we have
\begin{equation}
-\frac{1}{2 \pi}\; \frac{d{\alpha}_s}{ds}+\frac{1}{(2\pi)^2}\alpha_{s}^{2}\; 
\frac{dF^{(2)}_{s}(p,p')}{ds}=- \alpha_s^2 \; I_{s}^{(2)}(p,p') \; ,
\end{equation}
\noindent
where
\begin{eqnarray}
I_{s}^{(2)}(p,p')&=&\frac{1}{(2\pi)^2}\; \int_{0}^{\infty}dk \; k \; (2 
k^2-p^2-p'^2)\; e^{-2s[p^2 p'^2+k^4-k^2(p^2+p'^2)]}\nonumber\\
&=&\frac{1}{(2\pi)^2}\; \frac{e^{-2s \; p^2 p'^2}}{4s} \; .
\end{eqnarray}
\noindent
The equation for $\alpha_s$ is obtained by taking the limit $(p,p') \rightarrow 
0$, 
\begin{equation}
\frac{1}{2 \pi}\; \frac{d{\alpha}_s}{ds}= \alpha_s^2 \; I_{s}^{(2)}(0,0) \; ,
\label{alp22}
\end{equation}
\noindent
where
\begin{equation}
I_{s}^{(2)}(0,0)=\frac{1}{(2\pi)^2}\; \frac{1}{4s} \; .
\end{equation}
\noindent
Integrating Eq. (\ref{alp22}) from $s_0$ to $s$,
\begin{eqnarray}
&&\alpha_{s,2}=\frac{\alpha_{s_0}}{1- \frac{\alpha_{s_0}}{8 \pi}\;{\rm 
ln}\left(\frac{s}{s_0}\right)} \; .
\label{ralp22}
\end{eqnarray}
\noindent
In terms of the cutoff $\lambda$ we obtain
\begin{equation}
\alpha_{\lambda,2}=\frac{\alpha_{\lambda_0}}{1+\frac{\alpha_{\lambda_0}}{2 
\pi}\;{\rm ln}\left(\frac{\lambda}{\lambda_0}\right)} \; .
\end{equation}
\noindent
In principle,  knowing the value of $\alpha_{s_0}$ for a given $s_0$ we can 
determine the running coupling $\alpha_s$ for any $s$.  Since we cannot choose 
$s_0=0$ ($\lambda_0=\infty$) as in the $D=1$ case, to use Eq.~(\ref{ralp22}) we 
must specify a renormalization prescription that allows us to fix the coupling 
at some finite non-zero value of $s_0$. We discuss this issue in detail later 
in this subsection.

The equation for $F^{(2)}_{s}(p,p')$ is given by 
\begin{equation}
\frac{1}{(2\pi)^2}\frac{dF^{(2)}_{s}(p,p')}{ds}=I_{s}^{(2)}(0,0)- 
I_{s}^{(2)}(p,p') \; .
\end{equation}
\noindent
Integrating from $s_0$ to $s$ we obtain
\begin{eqnarray}
F^{(2)}_{s}(p,p')&=& \int_{s_0}^{s}ds' \; \frac{\left[1-e^{-2s' \; p^2 
p'^2}\right]}{4s'}\nonumber\\
&=& \frac{1}{4}\left[{\rm ln}\left(\frac{s}{s_0}\right)-{\rm Ei}(-2s \; p^2 \; 
p'^2)+{\rm Ei}(-2s_0 \; p^2 \; p'^2)\right]\; .
\end{eqnarray}
\noindent
Insisting that $F^{(2)}_{s}(p,p')=0$ when $p,p'=0$ we obtain
\begin{eqnarray}
F^{(2)}_{s}(p,p')&=& \frac{1}{4}\left[\gamma+{\rm ln}(2s \; p^2 \; p'^2)-{\rm 
Ei}(-2s \; p^2 \; p'^2)\right] \; .
\end{eqnarray}

At third-order we have
\begin{eqnarray}
-\frac{1}{2 \pi}\; \frac{d{\alpha}_s}{ds}+\frac{1}{(2\pi)^2}\; \alpha_{s}^{2}\; 
\frac{dF^{(2)}_{s}(p,p')}{ds}&+&\frac{2}{(2\pi)^2}\; \alpha_{s} \; 
\frac{d\alpha_{s}}{ds}\; F^{(2)}_{s}(p,p')+\frac{1}{(2\pi)^3}\; \alpha_{s}^{3} 
\; \frac{dF^{(3)}_{s}(p,p')}{ds}\nonumber\\
&=&- \alpha_s^2 \; I_{s}^{(2)}(p,p')+ \alpha_s^3 \; I_{s}^{(3)}(p,p') \; ,
\end{eqnarray}
\noindent
where
\begin{eqnarray}
I_{s}^{(3)}(p,p')&=&\frac{1}{(2\pi)^3}\;\int_{0}^{\infty}dk \; k \; (2 
k^2-p^2-p'^2)\; e^{-2s[p^2 p'^2+k^4-k^2(p^2+p'^2)]} \nonumber\\
&\times& \left[F^{(2)}_{s}(p,k)+F^{(2)}_{s}(k,p')\right] \; .
\end{eqnarray}
\noindent
In the limit $p,p' \rightarrow 0$ we obtain:
\begin{equation}
\frac{1}{2 \pi}\; \frac{d{\alpha}_s}{ds}= \alpha_s^2 \; I_{s}^{(2)}(0,0)- 
\alpha_s^3 \; I_{s}^{(3)}(0,0) \; ,
\label{alp32}
\end{equation}
\noindent
where
\begin{eqnarray}
I_{s}^{(3)}(0,0)=\frac{1}{(2\pi)^3}\; \int_{0}^{\infty}dk  \; 2 k^3 e^{-2s 
k^4}\; \left[F^{(2)}_{s}(0,k)+F^{(2)}_{s}(k,0)\right] \; .
\end{eqnarray}
\noindent
Since $ F^{(2)}_{s}(k,0)= F^{(2)}_{s}(0,k)=0$, the term proportional to 
$\alpha_s^3$ in Eq. (~\ref{alp32}) is zero.

The equation for $F^{(3)}_{s}(p,p')$ is given by 
\begin{equation}
\frac{1}{(2\pi)^3}\frac{dF^{(3)}_{s}(p,p')}{ds}=-\frac{1}{\pi} \; 
I_{s}^{(2)}(0,0)\; F^{(2)}_{s}(p,p')+ I_{s}^{(3)}(p,p') \; .
\end{equation}
\noindent
To obtain $F^{(3)}_{s}(p,p')$ the integrals in $k$ and $s$ must be evaluated 
numerically.

At fourth-order we obtain
\begin{eqnarray}
-\frac{1}{2 \pi}\; \frac{d{\alpha}_s}{ds}&+&\frac{1}{(2\pi)^2} \; 
\alpha_{s}^{2}\; \frac{dF^{(2)}_{s}(p,p')}{ds}
+\frac{2}{(2\pi)^2} \; \alpha_{s} \; \frac{d\alpha_{s}}{ds}\; 
F^{(2)}_{s}(p,p')\nonumber\\
&+&\frac{1}{(2\pi)^3}\; \alpha_{s}^{3} \; 
\frac{dF^{(3)}_{s}(p,p')}{ds}+\frac{3}{(2\pi)^3}\;\alpha_{s}^{2}\; 
\frac{d{\alpha}_s}{ds} \; F^{(3)}_{s}(p,p')
+\frac{1}{(2\pi)^4}\alpha_{s}^{4} \; \frac{dF^{(4)}_{s}(p,p')}{ds}\nonumber\\
&=&- \alpha_s^2 \; I_{s}^{(2)}(p,p')+ \alpha_s^3 \; I_{s}^{(3)}(p,p') + 
\alpha_s^4 \; I_{s}^{(4)}(p,p')  \; ,
\end{eqnarray}
\noindent
where
\begin{eqnarray}
I_{s}^{(4)}(p,p')&=&\frac{1}{(2\pi)^4}\;\int_{0}^{\infty}dk \; k \; (2 
k^2-p^2-p'^2)\; e^{-2s[p^2 p'^2+k^4-k^2(p^2+p'^2)]} \nonumber\\
&\times& \left[F^{(3)}_{s}(p,k)+F^{(3)}_{s}(k,p')+ F^{(2)}_{s}(p,k) 
F^{(2)}_{s}(k,p')\right] \; .
\end{eqnarray}
In the limit $p,p' \rightarrow 0$ we obtain:
\begin{equation}
\frac{1}{2 \pi}\; \frac{d{\alpha}_s}{ds}= \alpha_s^2 \; I_{s}^{(2)}(0,0) - 
\alpha_s^4 \; I_{s}^{(4)}(0,0) \; ,
\label{alp42}
\end{equation}
\noindent
where
\begin{eqnarray}
I_{s}^{(4)}(0,0)=\frac{1}{2\pi}\; \int_{0}^{\infty}dk \; k \; 2 k^2 e^{-2s 
k^4}\; \left[F^{(3)}_{s}(0,k)+F^{(3)}_{s}(k,0)\right] \; .
\end{eqnarray}
\noindent
For dimensional reasons Eq. (\ref{alp42}) takes the form
\begin{equation}
\frac{d{\alpha}_s}{ds}= \frac{B_2}{s} \; \alpha_s^2 - \frac{B_4}{s}\;  
\alpha_s^4 \; ,
\label{alp34}
\end{equation}
\noindent
where $B_2=\frac{1}{8\pi} $ and $B_4$ can be obtained by evaluating 
$I_{s}^{(4)}(0,0)$ (numerically) for $s=1$. 
\noindent
In terms of the cutoff $\lambda$ we obtain:
\begin{equation}
\frac{d{\alpha}_{\lambda}}{d\lambda}= \frac{1}{\lambda^{4}} \; B_2 \; 
\alpha_{\lambda}^{2} - \frac{1}{\lambda^4}\; B_4 \; \alpha_{\lambda}^{4} \; .
\end{equation}
\noindent
Integration of Eq. (\ref{alp42}) leads to a transcendental equation which is 
solved numerically in order to obtain the running coupling $\alpha_{s,4}$.

Qualitatively, the errors are expected to be a combination of inverse powers of 
$\lambda$ and powers (or inverse powers) 
of ${\rm ln}(\lambda)$ coming from the perturbative expansion in powers of 
$\alpha_{\lambda}$ for the coefficients of the irrelevant operators and the 
perturbative approximation for $\alpha_{\lambda}$. 

As pointed out above, to completely determine the renormalized hamiltonian at
a  given  order we need to specify the coupling at some scale $\lambda_0$. The
simplest  way is to choose a  value for the `exact' $\alpha_{\lambda_0}$.
Formally, this fixes the underlying  theory; {\it i.e.},  if we had the exact
hamiltonian (to all orders) we could obtain the exact values for all 
observables. However, since the hamiltonian is derived perturbatively and must 
be truncated  at some order in practical calculations, we can only obtain
approximate cutoff-dependent results for the observables. Moreover, in this
case the errors cannot be directly  evaluated, since the  exact values remain
unknown. As an example, we calculate the bound-state energy   choosing 
$\alpha_{\lambda_0}=1.45$ at $\lambda_0=100$. In Fig. 8 we show the 
binding-energy as a  function of the cutoff $\lambda$ using the following
approximations for the  interaction:
\vspace{0.5cm}

\noindent
(a) marginal operator with coupling ($\alpha_{\lambda_0}$),
\begin{equation}
V_{\lambda}(p,p')=-\frac{\alpha_{\lambda_0}}{2\pi}e^{-\frac{(p^2-p'^2)^2}
{\lambda^4}} \; ;
\end{equation}

\noindent
(b) marginal operator with running coupling renormalized to second-order 
($\alpha_{\lambda,2}$),
\begin{equation}
V_{\lambda}(p,p')=-\frac{\alpha_{\lambda,2}}{2\pi}\; 
e^{-\frac{(p^2-p'^2)^2}{\lambda^4}} \; ;
\end{equation}

\noindent
(c) marginal operator plus second-order irrelevant operator with running 
coupling renormalized to second-order ($\alpha_{\lambda,2}, 
F^{(2)}_{\lambda}$),
\begin{equation}
V_{\lambda}(p,p')=\left[-\frac{\alpha_{\lambda,2}}{2\pi}+\left(\frac{\alpha_{
\lambda,2}}{2\pi}\right)^2  \; F^{(2)}_{\lambda}(p,p')\right]\;  
e^{-\frac{(p^2-p'^2)^2}{\lambda^4}} \; ;
\end{equation}

\noindent
(d) marginal operator with running coupling renormalized to fourth-order  
($\alpha_{\lambda,4}$),
\begin{equation}
V_{\lambda}(p,p')=-\frac{\alpha_{\lambda,4}}{2\pi}\; 
e^{-\frac{(p^2-p'^2)^2}{\lambda^4}} \; ;
\end{equation}

\noindent
(e) marginal operator plus second-order irrelevant operator with running 
coupling renormalized to fourth-order $\alpha_{\lambda,4}, F^{(2)}_{\lambda}$),

\begin{equation}
V_{\lambda}(p,p')=\left[-\frac{\alpha_{\lambda,4}}{2\pi}+\left(\frac{\alpha_{
\lambda,4}}{2\pi}\right)^2  \; F^{(2)}_{\lambda}(p,p')\right] \; 
e^{-\frac{(p^2-p'^2)^2}{\lambda^4}} \; .
\end{equation}
\vspace{0.5cm}

\begin{figure}
\centerline{\epsffile{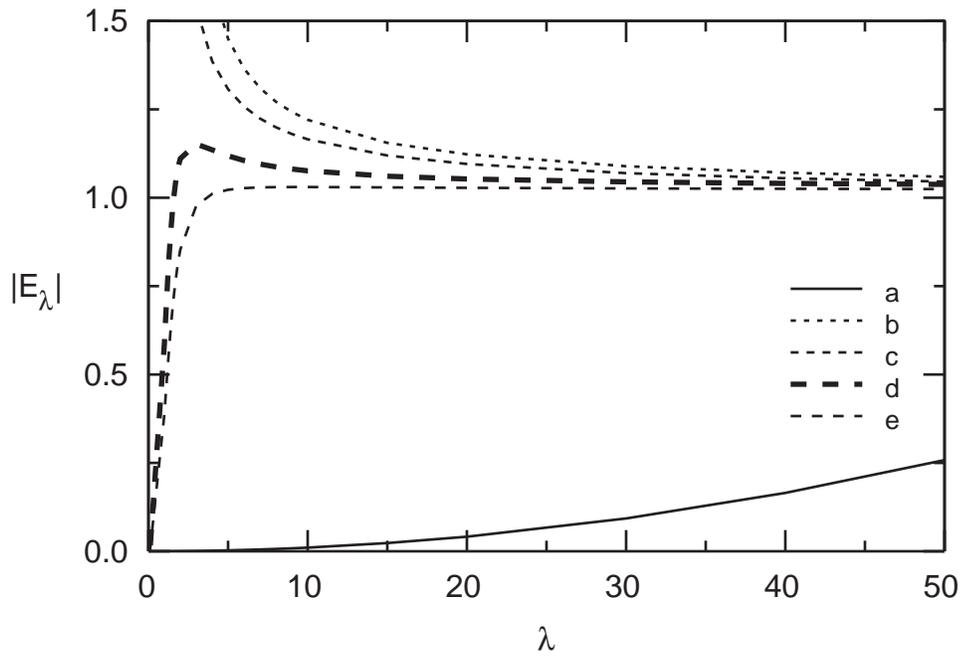}}
\vspace{0.5cm}
\caption{The binding energy for the two-dimensional delta-function potential 
with various approximations for the SRG hamiltonian. The exact theory is fixed 
by choosing $\alpha_0=1.45$ at $\lambda_0=100$.}
\end{figure}

We see that as the approximation is improved the cutoff dependence is reduced. 
As $\lambda \rightarrow \infty$ all curves should approach the same 
binding-energy, which corresponds to the exact value, and as $\lambda$ becomes 
small the perturbative approximation breaks down. 

A similar prescription is to find  $\alpha_{\lambda_0}$ at $\lambda_0$ that 
produces a given binding-energy, $E_0$. Since the fitting is implemented using 
a truncated hamiltonian, this $\alpha_{\lambda_0}$ is an approximation that 
becomes more accurate if we use a larger $\lambda_0$ and/or include higher 
order operators. Although in this case we can evaluate the errors, the scaling 
analysis becomes complicated as $\lambda \rightarrow \lambda_0$ because at this 
point we force the energy to be the exact value and so the error is zero. 

As an example, we calculate the bound-state energy when the coupling is fixed
at  $\lambda_0=100$ to give $E_0=1$. In Fig. 9 we show the `errors' in the
binding  energy using the same approximations listed above for the potential.
As  expected, all error lines drop abruptly to zero when $\lambda \rightarrow 
\lambda_0$, where the running coupling is chosen to fit what we define to be
the exact binding  energy. Away from this point we can analyze the errors.
With the hamiltonian  (a) (unrenormalized) we obtain a dominant error that
scales like ${\rm  ln}\left(\lambda_0/\lambda \right)$, corresponding to the
leading order error.  With the renormalized hamiltonian (b) the dominant
errors scale like $[{\rm  ln}\left(\lambda_0/\lambda \right)]^{-2}$, indicating
the elimination of the  leading order logarithmic errors. With the hamiltonian
(c) there is a small  shift, but no significant change in the error scaling.
The added irrelevant  operator may remove errors of order
$(\lambda_0/\lambda)^2$, but these are  smaller than the remaining 
$[{\rm ln}\left(\lambda_0/\lambda \right)]^{-2}$  errors. With hamiltonians (d)
and (e) in the range of intermediate cutoffs  ($E_0 << \lambda^2 <<
\lambda_0^2$) there is also only a shift in the errors.  The dips in (d) and
(e) correspond to values of $\lambda$ where the binding  energy equals the
exact value.

\begin{figure}
\centerline{\epsffile{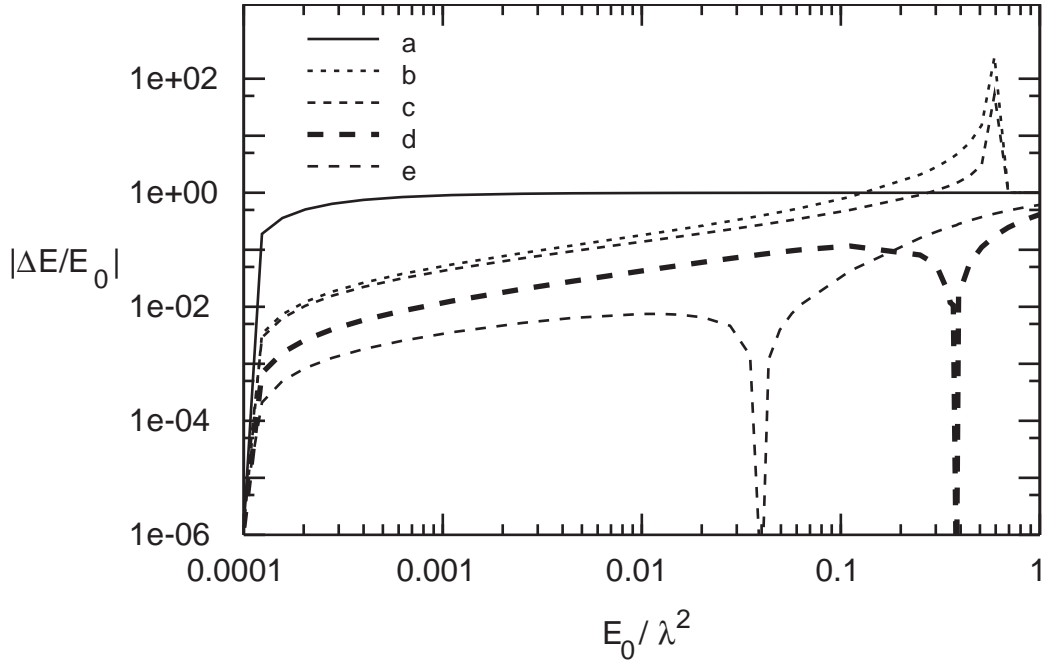}}
\vspace{0.5cm}
\caption{The SRG errors in the binding energy for the two-dimensional 
delta-function potential using various approximations for the SRG hamiltonian. 
The exact theory is fixed by choosing $E_0=1$ at $\lambda_0=100$.}
\end{figure}

This behavior is a perturbative artifact that 
can be understood in the following way. Consider the Schr\"{o}dinger equation 
with potential (d). Rescaling the momenta $p \rightarrow \lambda {\tilde p}$ we 
obtain
\begin{equation}
{\tilde p}^2 {\tilde \Phi}({\tilde p})-\frac{\alpha_{\lambda,4}}{2\pi}\; 
\int_{0}^{\infty} \; d{\tilde q} \; {\tilde q} \; e^{({\tilde p}^2-{\tilde 
q}^2)}\;{\tilde \Phi}({\tilde q})=\frac{E_{\lambda}}{\lambda^2}\; {\tilde 
\Phi}({\tilde p}) \; .
\end{equation}
\noindent
and
\begin{eqnarray}
E_{\lambda}=\lambda^2 \; \frac{\left[\int_{0}^{\infty}\; d{\tilde p} \; {\tilde 
p}\; \left({\tilde p}^2\;|{\tilde \Phi}({\tilde 
p})|^2\right)-\frac{\alpha_{\lambda,4}}{2\pi}\; \int_{0}^{\infty}\; d{\tilde p} 
\; {\tilde p}\; \int_{0}^{\infty} \; d{\tilde q} \; {\tilde q} \; e^{({\tilde 
p}^2-{\tilde q}^2)} \;{\tilde \Phi}({\tilde p})\;{\tilde \Phi}({\tilde 
q})\right]}{\int_{0}^{\infty}\; d{\tilde p} \; {\tilde p}\;|{\tilde 
\Phi}({\tilde p})|^2}\; .
\end{eqnarray}
As shown in Fig. 10, the coupling renormalized to fourth-order, 
$\alpha_{\lambda,4}$, approximately freezes for small $\lambda$ and as a 
consequence the bound-state energy scales like $E_{\lambda}\simeq \lambda^2 
\times {\rm constant}$ eventually becoming equal to the exact value and then 
deviating again. With the hamiltonian (e) the behavior is similar, with the dip 
occurring at a different value of $\lambda$ because of the irrelevant operator. 
As in the $D=1$ case, for small values of $\lambda$ the lines converge, 
indicating the breakdown of the perturbative expansion. 

\begin{figure}
\centerline{\epsffile{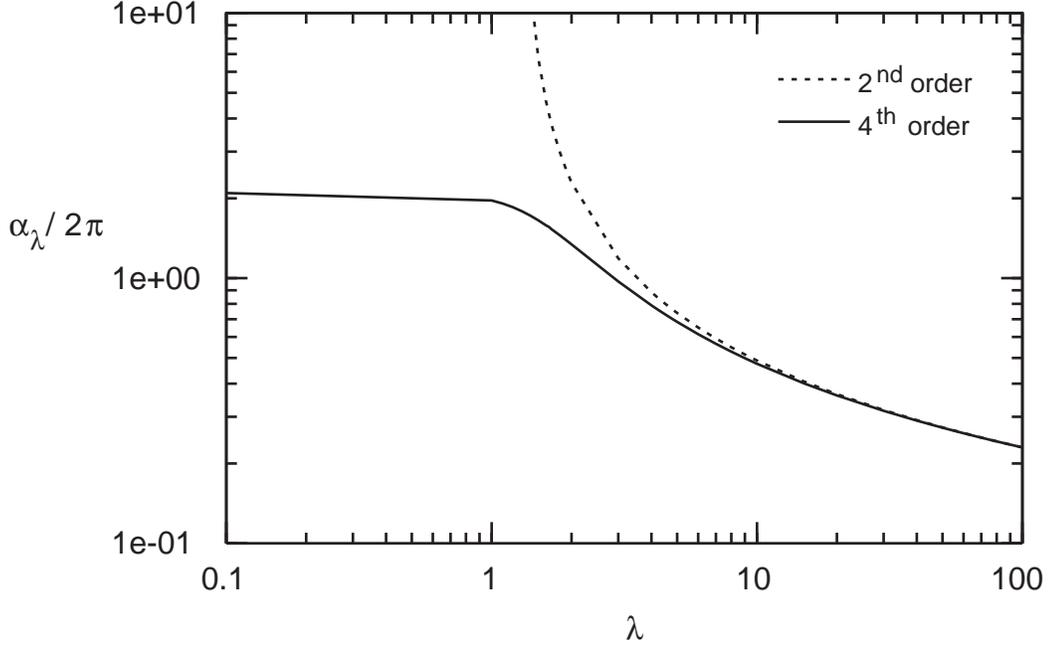}}
\vspace{0.5cm}
\caption{The SRG running coupling for the two-dimensional delta-function 
potential obtained with $\alpha_{\lambda_0}$ at $\lambda_0=100$ fixed to fit 
$E_0=1$.}
\end{figure}

An alternative prescription is to use the potential derived in subsection 5.1 
as the starting point for the similarity transformation. We introduce a large  
momentum cutoff $\Lambda$, define
\begin{equation}
\alpha_{s_0=0}=\alpha_{\Lambda}=\frac{4\pi}{{\rm 
ln}\left(1+\frac{\Lambda^2}{E_0}\right)} \; ,
\end{equation}   
\noindent
and set all irrelevant operators to zero at $s_0=0$. Note that the coupling 
$\alpha_{\lambda_0}$ is fixed at $\lambda_0=\infty$ by fitting the exact 
binding energy. With this definition the similarity hamiltonian with no
similarity cutoff  becomes well-defined and  we can set all of the similarity
transformation  boundary conditions at $s=0$. In some sense, the problem
becomes similar to the 
$D=1$ case. The previous derivation remains essentially the same. The only 
modification is that all integrals over momentum are cut off ($p \le  
\Lambda$).

At second-order we have
\begin{equation}
-\frac{1}{2 \pi}\; 
\frac{d{\alpha}_{s,\Lambda}}{ds}+\frac{1}{(2\pi)^2}\alpha_{s,\Lambda}^{2}\; 
\frac{dF^{(2)}_{s,\Lambda}(p,p')}{ds}=- \alpha_{s,\Lambda}^2 \; 
I_{s,\Lambda}^{(2)}(p,p')
\end{equation}
\noindent
where
\begin{eqnarray}
I_{s,\Lambda}^{(2)}(p,p')&=&\frac{1}{(2\pi)^2}\; \int_{0}^{\Lambda}dk \; k \; 
(2 k^2-p^2-p'^2)\; e^{-2s[p^2 p'^2+k^4-k^2(p^2+p'^2)]}\nonumber\\
&=&\frac{1}{(2\pi)^2}\; \frac{e^{-2s \; p^2 p'^2}}{4s} \; .
\end{eqnarray}
\noindent
The resulting second-order running coupling and irrelevant operator are given 
respectively by
\begin{equation}
\alpha_{s,\Lambda,2}=\frac{\alpha_{\Lambda}}{1-\frac{\alpha_{\Lambda}}{8\pi}\; 
\left[\gamma+{\rm ln}\left(2s \Lambda^4 \right)-{\rm Ei}\left(-2s \Lambda^4 
\right)\right]}
\end{equation}
\noindent
and
\begin{eqnarray}
F^{(2)}_{s,\Lambda}(p,p')&=&  \frac{1}{4}\left[\gamma+{\rm ln}(2s \; p^2 \; 
p'^2)-{\rm Ei}(-2s \; p^2 \; p'^2)\right]\nonumber\\
&+& \frac{1}{4}\left[\gamma+{\rm ln}(2s \Lambda^4)-{\rm Ei}(-2s 
\Lambda^4)\right]\nonumber\\
&-&\frac{1}{4}\left[\gamma+{\rm ln}\left(s \left[(p^2-\Lambda^2)^2 
+(p'^2-\Lambda^2)^2-(p^2-p'^2)^2\right]\right)\right.\nonumber\\
&&\left. \; \; \; \; \; -{\rm Ei}\left(-s \left[(p^2-\Lambda^2)^2 
+(p'^2-\Lambda^2)^2-(p^2-p'^2)^2\right]\right)\right]\; .
\end{eqnarray}

At third-order we have
\begin{eqnarray}
-\frac{1}{2 \pi}\; \frac{d{\alpha}_{s,\Lambda}}{ds}&+&\frac{1}{(2\pi)^2}\; 
\alpha_{s,\Lambda}^{2}\; 
\frac{dF^{(2)}_{s,\Lambda}(p,p')}{ds}+\frac{2}{(2\pi)^2}\; \alpha_{s,\Lambda} 
\; \frac{d\alpha_{s,\Lambda}}{ds}\; F^{(2)}_{s,\Lambda}(p,p')\nonumber\\
&+&\frac{1}{(2\pi)^3}\; \alpha_{s,\Lambda}^{3} \; 
\frac{dF^{(3)}_{s,\Lambda}(p,p')}{ds}=- \alpha_{s,\Lambda}^2 \; 
I_{s,\Lambda}^{(2)}(p,p')+ \alpha_{s,\Lambda}^3 \; I_{s,\Lambda}^{(3)}(p,p') \; 
,
\end{eqnarray}
\noindent
where
\begin{eqnarray}
I_{s,\Lambda}^{(3)}(p,p')&=&\frac{1}{(2\pi)^3}\;\int_{0}^{\Lambda}dk \; k \; (2 
k^2-p^2-p'^2)\; e^{-2s[p^2 p'^2+k^4-k^2(p^2+p'^2)]} \nonumber\\
&\times& \left[F^{(2)}_{s}(p,k)+F^{(2)}_{s}(k,p')\right] \; .
\end{eqnarray}
\noindent
In this case, 
\begin{eqnarray}
F^{(2)}_{s,\Lambda}(0,k)&=& \frac{1}{4}\left[\gamma+{\rm ln}(2s \Lambda^4)-{\rm 
Ei}(-2s \Lambda^4)\right]\nonumber\\
&=& \frac{1}{4}\left[\gamma+{\rm ln}(2s \Lambda^4-2s k^2 \Lambda^2)-{\rm 
Ei}(-2s \Lambda^4+2sk^2 \Lambda^2)\right]\; ,
\end{eqnarray}
\noindent
and so
\begin{eqnarray}
I_{s,\Lambda}^{(3)}(0,0)&=&\frac{1}{(2\pi)^3}\; \frac{1}{4}\; 
\int_{0}^{\Lambda}dk  \; 4 k^3 e^{-2s k^4}\; \left(\left[\gamma+{\rm ln}(2s 
\Lambda^4)-{\rm Ei}(-2s \Lambda^4)\right]\right. \nonumber\\
&-&\left. \left[\gamma+{\rm ln}(2s \Lambda^4-2s k^2 \Lambda^2)-{\rm Ei}(-2s 
\Lambda^4+2sk^2 \Lambda^2)\right]\right)\; .
\end{eqnarray}
\noindent
Since $I_{s,\Lambda}^{(3)}(0,0) \neq 0$ the term proportional to 
$\alpha_{s,\Lambda}^3$ in Eq. (\ref{alp32}) does not vanish. To obtain 
$\alpha_{s,\Lambda,3}$ we evaluate $I_{s,\Lambda}^{(3)}(0,0)$ and solve 
Eq. (\ref{alp32}) numerically.

The equation for $F^{(3)}_{s,\Lambda}(p,p')$ is given by 
\begin{equation}
\frac{1}{(2\pi)^3}\frac{dF^{(3)}_{s,\Lambda}(p,p')}{ds}=-\frac{1}{\pi} \; 
I_{s,\Lambda}^{(2)}(0,0)\; F^{(2)}_{s,\lambda}(p,p')+ I_{s,\Lambda}^{(3)}(p,p') 
\; .
\end{equation}
\noindent
To obtain $F^{(3)}_{s,\Lambda}(p,p')$ the integrals over $k$ and $s$ must be 
evaluated numerically. In the limit $s\Lambda^4 \rightarrow \infty$ with $s$ 
fixed at some non-zero value
\begin{eqnarray}
&&F^{(2)}_{s,\Lambda}(p,p')\rightarrow \frac{1}{4}\left[\gamma+{\rm ln}(2s \; 
p^2 \; p'^2)-{\rm Ei}(-2s \; p^2 \; p'^2)\right] \; , \\
&&I_{s,\Lambda}^{(3)}(0,0) \rightarrow 0 \; ,
\end{eqnarray}
\noindent
and $\alpha_{s,\Lambda}$ becomes indeterminate, requiring the coupling to be 
fixed at some $s_0 \neq 0$. In this way, we recover the result of the previous 
prescription.

Although less trivial, this prescription allows a more transparent error 
analysis. We can also extend the calculation to larger values of the similarity 
cutoff, $\lambda$, and analyze the errors in a different scaling regime. In 
Fig. 11 we show the errors in the binding-energy  obtained using the following 
approximations for the potential with $\Lambda=50$:
\vspace{0.5cm}

\noindent
(a) marginal operator with coupling ($\alpha_{\lambda_0}$),
\begin{equation}
V_{\lambda,\Lambda}(p,p')=-\frac{\alpha_{\Lambda}}{2\pi}\; 
e^{-\frac{(p^2-p'^2)^2}{\lambda^4}} \; ;
\end{equation}

\noindent
(b) marginal operator with running coupling renormalized to second-order 
($\alpha_{\lambda,\Lambda,2}$),
\begin{equation}
V_{\lambda,\Lambda}(p,p')=-\frac{\alpha_{\lambda,\Lambda,2}}{2\pi}\; 
e^{-\frac{(p^2-p'^2)^2}{\lambda^4}} \; ;
\end{equation}

\noindent
(c) marginal operator plus second-order irrelevant operator with running 
coupling renormalized to second-order ($\alpha_{\lambda,\Lambda,2}, 
F^{(2)}_{\lambda}$),
\begin{equation}
V_{\lambda,\Lambda}(p,p')=\left[-\frac{\alpha_{\lambda,\Lambda,2}}{2\pi}+\left(
\frac{\alpha_{\lambda,\Lambda,2}}{2\pi}\right)^2 \; 
F^{(2)}_{\lambda,\Lambda}(p,p')\right] \; e^{-\frac{(p^2-p'^2)^2}{\lambda^4}} 
\; ;
\end{equation}

\noindent
(d) marginal operator with running coupling renormalized to third-order  
($\alpha_{\lambda,\Lambda,3}$),
\begin{equation}
V_{\lambda,\Lambda}(p,p')=-\frac{\alpha_{\lambda,\Lambda,3}}{2\pi}\; 
e^{-\frac{(p^2-p'^2)^2}{\lambda^4}} \; ;
\end{equation}

\noindent
(e) marginal operator plus second-order irrelevant operator with running 
coupling renormalized to third-order $\alpha_{\lambda,\Lambda,3}, 
F^{(2)}_{\lambda}$),
\begin{equation}
V_{\lambda,\Lambda}(p,p')=\left[-\frac{\alpha_{\lambda,\Lambda,3}}{2\pi} 
+\left(\frac{\alpha_{\lambda,\Lambda,3}}{2\pi}\right)^2 \; 
F^{(2)}_{\lambda,\Lambda}(p,p')\right] \; e^{-\frac{(p^2-p'^2)^2}{\lambda^4}}  
\; .
\end{equation}
\vspace{0.5cm}

\begin{figure}
\centerline{\epsffile{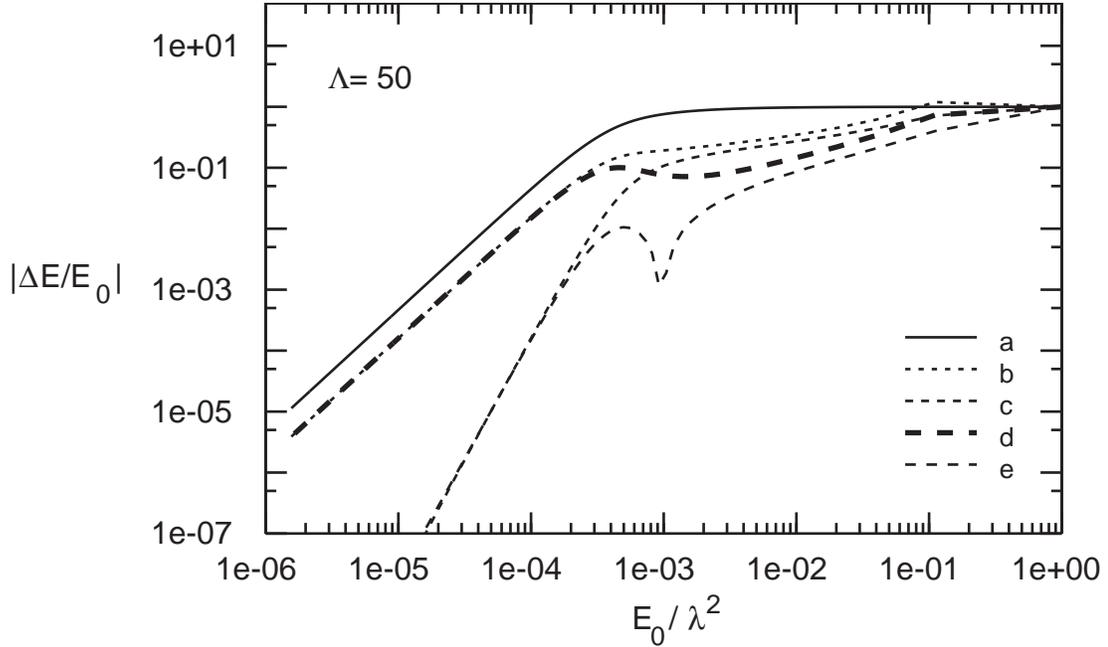}}
\vspace{0.5cm}
\caption{The SRG errors in the binding energy for the two-dimensional 
delta-function potential using various approximations for the similarity 
hamiltonian. The exact theory is fixed by regulating the ``bare hamiltonian'' 
using a sharp momentum cutoff, $\Lambda$, and letting the bare coupling depend 
on $\Lambda$ such that the binding energy is fixed. We use $\Lambda=50$ and 
$E_0=1$.}
\end{figure}

There are clearly two distinct scaling regions when an additional large
momentum cutoff $\Lambda$ is placed on the initial matrix and a similarity
cutoff is then applied. When the similarity cutoff is larger than
$\Lambda$, we see scaling behavior that is similar to EFT. Curves (a), (b),
and (d) all have the same slope. None of these hamiltonians contains
irrelevant operators, but the marginal coupling differs in each. All results
become exact as the similarity cutoff goes to infinity, and these curves are
close to one another because the coupling runs little in this region. Curves
(c) and (e) show that there is a power-law improvement when irrelevant
operators are added, and that once again when the similarity cutoff is larger
than $\Lambda$ an improvement in the running coupling makes little difference.
These results closely resemble those in EFT and we conclude that it is the
leading irrelevant operator that leads to this improvement. Even though the
coupling in front of this operator is approximated by the first term in an
expansion in powers of the running coupling, the coupling is sufficiently
small that this approximation works well and the operator eliminates most of
the leading power-law error in curves (a), (b), and (d).

When the similarity cutoff become smaller than $\Lambda$ we see a crossover to
a more complicated scaling regime that resembles the SRG scaling discussed
above.  The error displayed by curve (a) approaches 100\%, while the running
coupling introduced in curve (b) reduces the error to an inverse logarithm.
Improving the running coupling in curve (d) further reduces the error, and we
see that curve (c) crosses curve (d) at a point where improving the running
coupling becomes more important than adding irrelevant operators. As above,
the best results require us to both improve the running coupling by adding
third-order corrections and add the second-order irrelevant operators. In no
case do we achieve power-law improvement, because as we have discussed there
are always residual inverse logarithmic errors. Had we fit the running
coupling to data, as we would do in a realistic calculation, we would obtain
power-law improvement and the residual error would be proportional to an
inverse power of the cutoff times an inverse power of the logarithm of the
cutoff. 

In Fig. 12 we show the running coupling at 2nd and 3rd order. Although the 3rd
order corrections are small for all $\lambda$ and vanish when $\Lambda
\rightarrow \infty$, the improvement resulting from this correction in Fig. 11
is significant.

\begin{figure}
\centerline{\epsffile{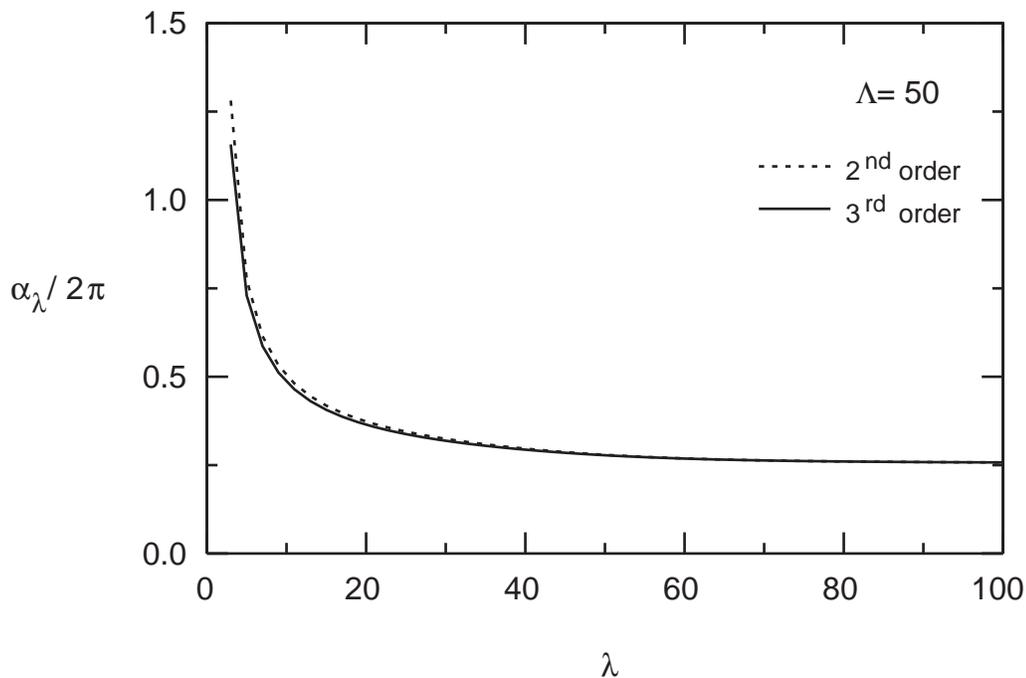}}
\vspace{0.5cm}
\caption{The SRG running coupling for the two-dimensional delta-function 
potential renormalized to second and third-order obtained with 
$\alpha_{\lambda_0=\infty}\rightarrow \alpha_{\Lambda}=4\pi/{\rm 
ln}\left(1+\frac{\Lambda^2}{E_0}\right)$.}
\end{figure}

The scaling behavior with a large momentum cutoff $\Lambda$ in place is
complicated, but it is fairly straightforward to understand it and to find a
sequence of approximations that systematically improve the non-perturbative
results. The calculations become increasingly complicated, but at each order
one must improve the running coupling, or fit it to data, and add higher order
irrelevant operators. In a field theory we need to let $\Lambda \rightarrow
\infty$ and study the scaling behavior of the theory in the regime where
$\lambda \ll \Lambda$. Although we do not display a compete set of figures, in
Fig. 13 we show what happens to the running coupling as $\Lambda$ is
increased, with the bound state energy fixed at one.

\begin{figure}
\centerline{\epsffile{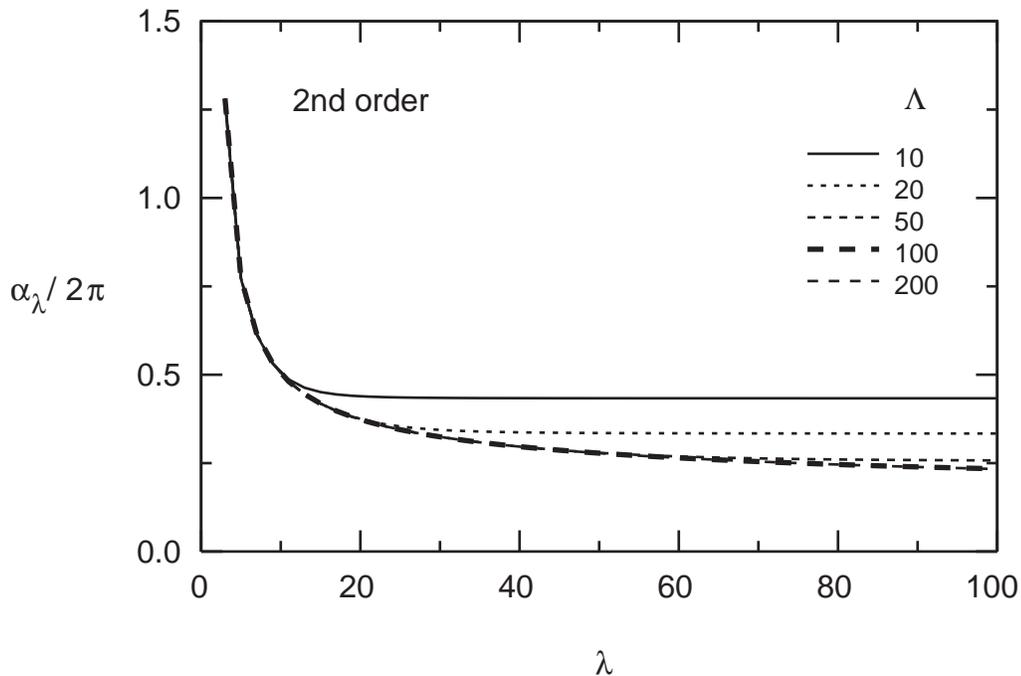}}
\vspace{0.5cm}
\caption{The SRG running coupling for the two-dimensional delta-function 
potential renormalized to second-order obtained with 
$\alpha_{\lambda_0=\infty}\rightarrow \alpha_{\Lambda}=4\pi/{\rm 
ln}\left(1+\frac{\Lambda^2}{E_0}\right)$.}
\end{figure}

As is evident in the exact solution, as $\Lambda$ increases the coupling
decreases. When $\lambda \gg \Lambda$, the coupling runs slowly and stays near
its asymptotic value. As $\lambda$ approaches $\Lambda$ the coupling begins to
run noticeably, and when $\lambda$ becomes much less than $\Lambda$ the
coupling approaches a universal curve that is insensitive to its asymptotic
value. Plots of the error in the binding energy for various approximations and
different values of $\Lambda$ closely resemble Fig. 11, with two scaling
regimes whose boundary is $\lambda = \Lambda$.

We close this section by reminding the reader that in all of these
calculations there is only one free parameter. In a realistic calculation we
would fit this parameter to a binding energy and we would expect to see
residual errors in other observables that is inversely proportional to powers
of the cutoff and logarithms of the cutoff.
%

%

%

\section{Conclusions}

We have illustrated the EFT and SRG methods for producing effective cutoff
hamiltonians using the one-dimensional and two-dimensional delta-function
potentials. We have shown that the addition of irrelevant operators with
couplings tuned to fit low energy scattering observables leads to errors that
scale as inverse powers of the cutoff in EFT. We have shown that the SRG with
coupling coherence leads to errors that scale as inverse powers of the cutoff
and logarithms of the cutoff. EFT typically leads to smaller errors than the
SRG, but the cost is an increased number of parameters that must be fit to
data. The SRG with coupling coherence requires the same number of parameters
as the underlying `fundamental' theory, but the cost is exponentially
increasing algebraic complexity to remove errors that contain inverse powers
of logarithms of the cutoff.

Our examples do not adequately illustrate the real utility of these
renormalization tools, because our examples are easily solved with other
methods. EFT can be used when the underlying theory cannot be solved and even
when it is not known. All that is required is a knowledge of the degrees of
freedom and symmetries that appear at `low' energies. The SRG can also be
employed when the underlying theory is not known, but in general it offers no
advantages over EFT unless that effective theory is valid over very many energy
scales. When the effective theory is the same as the fundamental theory, but
with cutoffs introduced, or when it applies over very many scales, it may be
possible to fix some of the couplings in terms of the fundamental couplings
without using extra data.

%
\section{Acknowledgments}

We would like to acknowledge many useful discussions with Brent Allen,
Martina Brisudova, Dick Furnstahl, Stan G{\l}azek, Billy Jones, Roger Kylin,
Rick Mohr, Jim Steele, and Ken Wilson. This work was supported by National
Science Foundation grant PHY-9800964, and S.S. was supported by a CNPq-Brazil
fellowship (proc. 204790/88-3).
%

%

\newpage
%

%

\centerline{\LARGE \bf References}
\def\refname{}

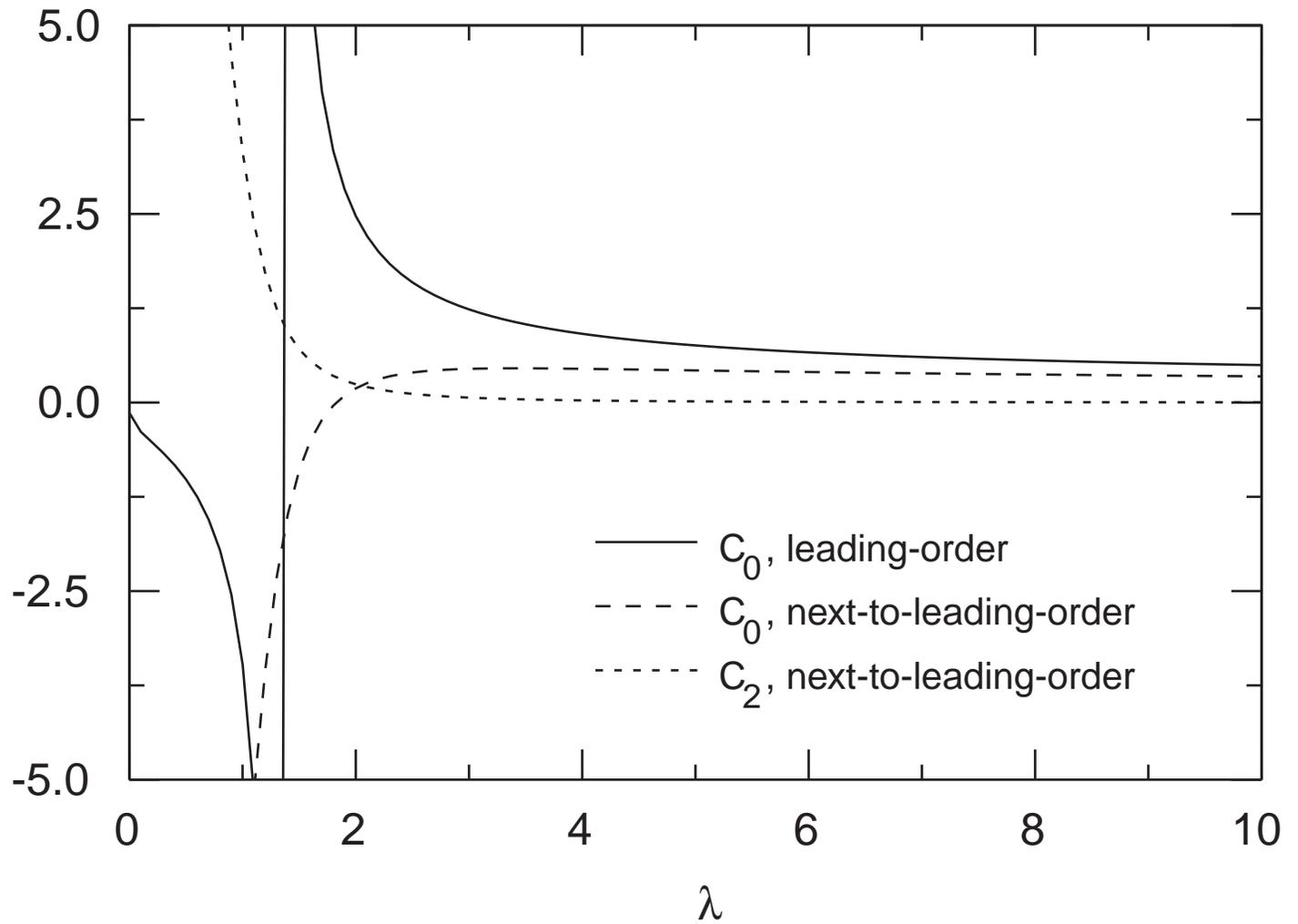
\end{document}